\documentclass[useAMS,usenatbib]{mn2e}

\usepackage{color}
\usepackage{lscape,graphicx}
\usepackage{amsmath}
\usepackage{journal_shortcuts}
\usepackage{amssymb}
\usepackage{multirow}
\usepackage{afterpage}
\usepackage{ulem}
\def\cm{{\rm\thinspace cm}}

\def\erg{{\rm\thinspace erg}}

\def\ga{{\rm\thinspace gauss}}

\def\s{{\rm\thinspace s}}

\def\ergpscmps{\hbox{$\erg\cm^{-2}\s^{-1}\,$}}

\def\ciza{\hbox{CIZA J2242.8+5301}}

\def\h0{\hbox{{\rm H}$^0$}}
\DeclareMathAlphabet{\vib}{OML}{cmm}{m}{it}
\newcommand*{\satellite}[1]{\textit{#1}}
\newcommand*{\xmm}{\satellite{XMM-Newton}}

\newcommand*{\rosat}{\satellite{ROSAT}}

\newcommand{\lsim}{\mathrel{\hbox{\rlap{\lower.55ex\hbox{$\sim$}} \kern-.3em \raise.4ex \hbox{$<$}}}}
\newcommand{\gsim}{\mathrel{\hbox{\rlap{\lower.55ex\hbox{$\sim$}} \kern-.3em \raise.4ex \hbox{$>$}}}}

\title[XMM observations of the merging cluster CIZA J2242]{\xmm\ observations of the merging galaxy cluster CIZA~J2242.8+5301}
\author[G.~A.~Ogrean et al.]{G.~A.~Ogrean$^{1}$\thanks{E-mail:
gogrean@hs.uni-hamburg.de}, M.~Br\"{u}ggen$^{1}$,  H.~R\"{o}ttgering$^{2}$, A. Simionescu$^{3}$, J.~H.~Croston$^{4}$, \and R. van Weeren$^{2}$, M. Hoeft$^{5}$\\
$^{1}$Hamburger Sternwarte, Gojenbergsweg 112, 21029 Hamburg, Germany\\
$^{2}$Leiden Observatory, Leiden University, P.O. Box 9513, NL-2300 RA Leiden, Netherlands\\
$^{3}$KIPAC, Stanford University, 452 Lomita Mall, Stanford, CA 94305, USA\\
$^{4}$School of Physics and Astronomy, University of Southampton, Southampton, SO17 1SJ\\
$^{5}$Th\"uringer Landessternwarte Tautenburg, Sternwarte 5, 07778 Tautenburg, Germany}

\begin{document}

\date{Accepted xxx xxxx xx. Received xxx xxxx xx; in original form xxx xxxx xx}

\pagerange{\pageref{firstpage}--\pageref{lastpage}} \pubyear{2012}

\maketitle

\label{firstpage}

\begin{abstract}
We studied the intracluster medium of the galaxy cluster CIZA J2242.8+5301 using deep \emph{XMM-Newton} observations. The cluster hosts a remarkable 2-Mpc long, $\sim 50$-kpc wide radio relic that has been nicknamed the ``Sausage''. A smaller, more irregular counter-relic is also present, along with a faint giant radio halo. We analysed the distribution of the ICM physical properties, and searched for shocks by trying to identify density and temperature discontinuities. East of the southern relic, we find evidence of shock compression corresponding to a Mach number of $\sim 1.3$, and speculate that the shock extends beyond the length of the radio structure. The ICM temperature increases at the northern relic. More puzzling, we find a ``wall'' of hot gas east of the cluster centre. A partial elliptical ring of hot plasma appears to be present around the merger. While radio observations and numerical simulations predict a simple merger geometry, the X-ray results point towards a more complex merger scenario.
\end{abstract}


\section{Introduction}
\label{s:intro}

Galaxy clusters form hierarchically, via mergers with other clusters and groups of galaxies, and also by accretion of gas from the intergalactic medium (IGM). Mergers may trigger shocks and turbulence in the intracluster medium (ICM), and extended sources of diffuse synchrotron emission are typically observed in clusters that show evidence for a merger \citep[e.g.][]{buote2001, govoni2004, Venturi2008, cassano10, vanWeeren2011a}.

It has been proposed that relics trace shock waves injected into the ICM during a major merger. The commonly accepted theory is that in the presence of a magnetic field, particles are (re-)accelerated at the shock to relativistic energies and emit synchrotron radiation. This emission is visible in the radio as arc-shaped relics at the cluster periphery. 

Radio observations reveal the position of the relic, while X-ray observations can locate the shock wave by identifying surface brightness discontinuities, or temperature, pressure, and entropy changes within the ICM. Therefore, multiwavelength observations of merging galaxy clusters harboring relics are necessary in order to test the radio relic-shock wave hypothesis. Moreover, such observations provide information about the merger dynamics, the processes behind relic formation, and the intracluster magnetic fields in the region of the relic. At the same time, they allow us to refine models of shock acceleration in merging clusters, where the low Mach numbers suggest an efficiency inadequate to explain the existence of radio relics by direct shock acceleration \citep[e.g.,][]{Kang2007}.

So far, only a handful of shock fronts have been found that exhibit, both, a sharp gas density edge and an unambiguous temperature jump: in the "Bullet Cluster", 1E065756 \citep{markevitch02}, Abell 520 \citep{markevitch05}, Abell 2146 \citep{russell10}, Abell 3667 \citep{finoguenov10}, and Abell 754 \citep{Macario2011}. Of these, only the shocks in Abell 3667 and Abell 754 are associated with radio relics, even though there are about 50 known relics. Such discoveries are rare because the merger has to be observed at the time when the shock has not yet moved to the low-brightness outskirts and is propagating nearly in the plane of the sky in order to give a clear view of the shock discontinuity.

Here, we present a combined X-ray and radio analysis of the merging galaxy cluster CIZA J2242.8+5301 ($z=0.1921$), an X-ray luminous cluster ($L_{\rm X}=6.8\times 10^{44}$ erg/s in the \emph{ROSAT} band $0.1-2.4$ keV) which hosts a double radio relic, and an extended, faint radio halo. The properties of the relics have been analysed in detail by \citet{vanWeeren2010}; in this introduction, we present a summary of their findings. The relics are oriented along the N-S direction. The northern relic is the most spectacular, with a length of 2 Mpc, and a width of only 55~kpc, narrower than any other known relic. At this relic, deep radio observations have revealed a spectral index gradient towards the centre, with the spectral index changing from $-0.6$ to $-2.0$ ($\mathcal{F}_{\nu}\propto \nu^{-\alpha}$, where $\mathcal{F}_{\nu}$ is the radio flux at frequency $\nu$, and $\alpha$ is the spectral index). Furthermore, the relic is strongly polarized, at a level of $\sim 50\%$, with magnetic field vectors aligned with the relic (see fig. 3 in \citet{vanWeeren2010}). These radio properties provide clear evidence for shock acceleration and spectral ageing at an outward moving shock.

\citet{vanWeeren2011b} used hydrodynamical simulations of binary cluster mergers to constrain the merger geometry in CIZA J2242.8+5301, based on the observed radio structures. They found that the cluster is most likely undergoing a merger in the plane of the sky, between two clusters of nearly equal mass (mass ratio of about 2:1) that collided with a small impact parameter ($\lesssim 400$ kpc). In the following, we explore the spatial and spectral properties of the ICM using a deep \xmm\ observation of the cluster, and we compare the predicted merger scenario with that implied by the X-ray results, in order to further refine our understanding of the merger geometry that led to the formation of the spectacular double-relic system.

The paper is organized as follows: Section \ref{s:obs} presents the observations and the data reduction. Section \ref{s:analysis} details our analysis, while in Section \ref{s:syserrors} we analyse the effect of systematic uncertainties on the spectral measurements. Sections \ref{s:shocks}, \ref{s:smudge}, and \ref{s:mergerscenario} discuss the results. A summary of our findings is given in Section \ref{s:conclusions}.

We assume a flat $\Lambda$CDM universe with $H_0=70$ km\,s$^{-1}$\,Mpc$^{-1}$, $\Omega_{\rm M}=0.3$, and $\Omega_{\rm \Lambda}=0.7$. At the redshift of the cluster, 1 arcmin corresponds to 192 kpc. Throughout the paper, quoted errors are $1\sigma$ statistical errors, unless stated otherwise. Normalizations of all the spectral components are given in default {\sc Xspec} units (i.e., cm$^{-5}$ for APEC components, and photons\,keV$^{-1}$\,cm$^{-2}$\,s$^{-1}$, measured at 1 keV, for power-law components) per square arcmin. Temperatures are in units of keV. To enhance the clarity of the spectral plots, all spectra have been rebinned in {\sc Xspec} to a minimum significance of $3\sigma$, while setting a limit of 30 for the maximum number of combined adjancent bins.

\section{Observations and data reduction}
\label{s:obs}

\subsection{Radio}
\label{s:radio}

CIZA J2242.8+5301 was observed with the Westerbork Synthesis Radio Telescope (WSRT) in the  21~cm band. The observations were spread out between March and November 2009. The total time on target time was about 30~hrs. A bandwidth 160~MHz of was used, evenly divided over 8 sub-bands. All four linear polarization products were recored with 64 channels per sub-band. The data were calibrated using the CASA\footnote{http://casa.nrao.edu/} package.

The reduction consisted of standard gain and bandpass calibration using a flux calibrator observed at the start and end of a run. Time ranges for antennas affected by shadowing were removed and some data were flagged due to radio frequency interference. The fluxes for the calibrator sources were set according to the \cite{perleyandtaylor} extension to the \cite{1977A&A....61...99B} scale. Several rounds of phase self-calibration, followed by two rounds of amplitude and phase self-calibration were carried out. An image with robust weighting \citep{briggs_phd} of 0 was made. The image was cleaned using manually placed clean boxes. The resolution of this image is $16.5\arcsec \times 12.9\arcsec$ and the noise is 19~$\mu$Jy~beam$^{-1}$.

\subsection{X-ray}
\label{s:xray}

\begin{figure}
 \begin{center}
  \includegraphics[width=\columnwidth,keepaspectratio=true,clip=true,trim=0.2cm 0cm 1.3cm 0cm]{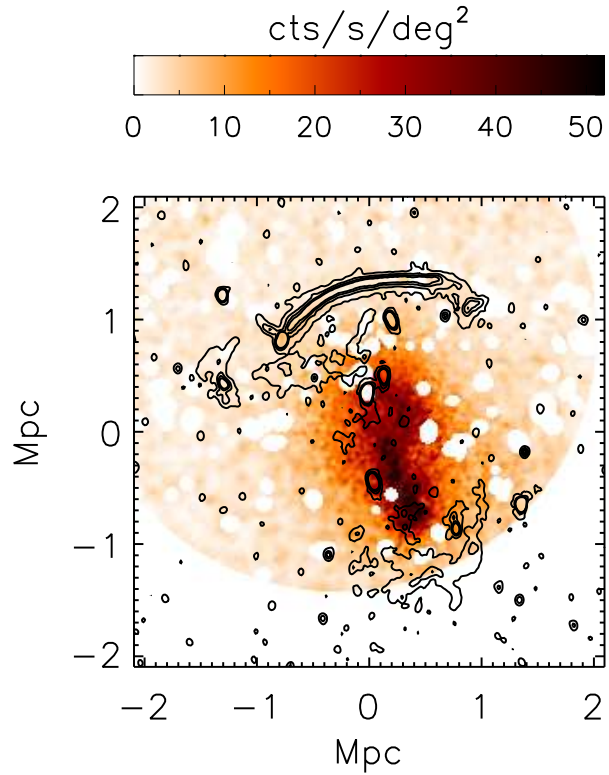}
 \end{center}
 \caption{Exposure-corrected and instrumental background-subtracted \emph{XMM-Newton} EPIC surface brightness map of the cluster, in the energy band $0.5-4$ keV. The image was binned by 2, and adaptively smoothed with a minimum of 100 counts per bin. Overlaid are WSRT 1.4 GHz radio contours, drawn at $[1,5,10,...]\times 100$ $\mu$Jy/beam.}
 \label{fig:xmm-radio}
\end{figure}

CIZA J2242.8+5301 was observed on December 13-15, 2010 for 127 ks with the three EPIC cameras of \emph{XMM-Newton}. The observations were performed in full-frame mode, using the medium filter.

Data reduction for all cameras used the Extended Source Analysis Software ({\sc esas}) integrated in the \emph{XMM-Newton} Science Analysis System ({\sc sas}) version 12.0.1. Raw event files were created from the Observation Data Files (ODF) using the {\sc xmm-esas} routines \texttt{emchain} and \texttt{epchain}. As \xmm\ orbits the Earth, protons with energies less than a few hundred keV can interact with the instruments, and be funneled towards the detectors. These soft protons (SPs) will create a time-variable (flaring) instrumental background component inside the field of view (FOV). The data was filtered of SP flares using the \texttt{mos-filter} and \texttt{pn-filter} routines, which apply a $1.5\sigma$ cutoff around the center of a Gaussian fitted to the EPIC lightcurve histograms. We then checked each filtered event file for residual soft proton contamination. This was done by comparing the EPIC count rates inside the field of view (thus, affected by possible residual soft protons) to the corresponding count rates in the unexposed corners of the detectors. The count rates were calculated in the energy bands $10-12$ keV (MOS) and $12-14$ keV (pn), in which the detectors are essentially insensitive to X-ray photons. The higher the count rate ratio inside-to-outside the FOV ($\mathcal{R}_{\rm SP}$) in these energy bands, the higher the level of residual SP contamination. \citet{delucamolendi2004} found that count rate ratios below 1.15 indicate event files that are hardly contaminated by residual soft protons. According to the same authors, count rate ratios between 1.15 and 1.3 correspond to event files slightly contaminated by soft protons, while ratios above 1.3 signify that the event files are significantly contaminated by flares. The values of $\mathcal{R}_{\rm SP}$ following the \texttt{mos-filter} run are shown in Table \ref{tab:spresiduals}. For the pn data, the flare-filtered event file created by the \texttt{pn-filter} routine had $\mathcal{R}_{\rm SP} = 1.27\pm 0.049$. Therefore, we created pn light curves and count rate histograms in the energy bands $2-5$ and $12-14$ keV (most sensitive to flares), and gradually lowered the count rate threshold of good-time intervals down to the maximum level at which $\mathcal{R}_{\rm SP}$ was below 1.15. After filtering, the total good-time intervals were 60 ks for MOS1, 62 ks for MOS2, and 25 ks for pn. Table \ref{tab:spresiduals} lists the $\mathcal{R}_{\rm SP}$ values for the three filtered EPIC event files.

\begin{table}
\caption{Residual soft proton contamination for the MOS and pn event files, based on the $\mathcal{R}_{\rm SP}$ parameter. We list the values calculated in the hard band, i.e. $10-12$ keV for MOS, and $12-14$ keV for pn. Following \citet{delucamolendi2004} and \citet{LeccardiMolendi2008}, we also calculated $\mathcal{R}_{\rm SP}$ in the energy bands $8-12$ and $6-12$ keV, respectively, for the MOS event files.}
\label{tab:spresiduals}
\footnotesize{
\begin{tabular}{cccc}
 \hline
  Energy band & MOS 1 & MOS 2 & pn \\ 
 \hline
   $6-12$ keV & $1.10\pm 0.024$ & $1.07\pm 0.022$ & --                \\
   $8-12$ keV & $1.09\pm 0.028$ & $1.10\pm 0.026$ & --                \\
  $10-12$ keV & $1.10\pm 0.037$ & $1.08\pm 0.033$ & --                \\
  $12-14$ keV & --               & --               & $1.10\pm 0.044$ \\
 \hline
\end{tabular}
}
\end{table}

Contaminating point-like sources within the FOV were identified with the {\sc xmm-esas} routine \texttt{cheese}, using a flux threshold of $3\times 10^{-15}$ \ergpscmps. The output files were then checked for spurious detections and undetected sources. All point-like sources were excluded from the analysis.

Spectra and images were extracted using the routines \texttt{mos-spectra} and \texttt{pn-spectra}. The resulting data, along with the {\sc xmm-esas} CalDB files\footnote{ftp://xmm.esac.esa.int/pub/ccf/constituents/extras/esas\_caldb/} describing the quiescent particle background, were then used by \texttt{mos\_back} and \texttt{pn\_back} to create images and spectra of the quiescent particle background (QPB). We used the \xmm\ CalDB files released in June 2012.

The pn data was corrected for out-of-time (OoT) events. OoT events are recorded during the readout time of the CCDs, and are therefore assigned an incorrect RAWY position. For the full-frame mode, the ratio of the readout time to the total integration time is 6.3\% for the pn detector (only 0.35\% for MOS, for which no correction is necessary). Therefore, the {\sc xmm-esas} routines create spectra and images of the OoT events, scale them by 6.3\%, and subtract them from the pn images and spectra.

CCD \#6 of MOS1 became unoperational in 2005, and is automatically excluded by {\sc xmm-esas}.

% \begin{figure}
%  \begin{center}
%   \includegraphics[width=\columnwidth,keepaspectratio=true,clip=true,trim=0.5cm 0cm 1cm 0cm]{radiorosat.ps}
%  \end{center}
%  \caption{1.4 GHz WSRT image of the double-relic system. Equidistant contours correspond to the \emph{ROSAT} X-ray emission from the cluster's ICM.}
%  \label{fig:rosat}
% \end{figure}

\begin{figure*}
 \begin{center}
  \vspace{-1.0cm}
  \includegraphics[width=\textwidth,keepaspectratio=true,clip=true,trim=0.5cm 0cm 1cm 3cm]{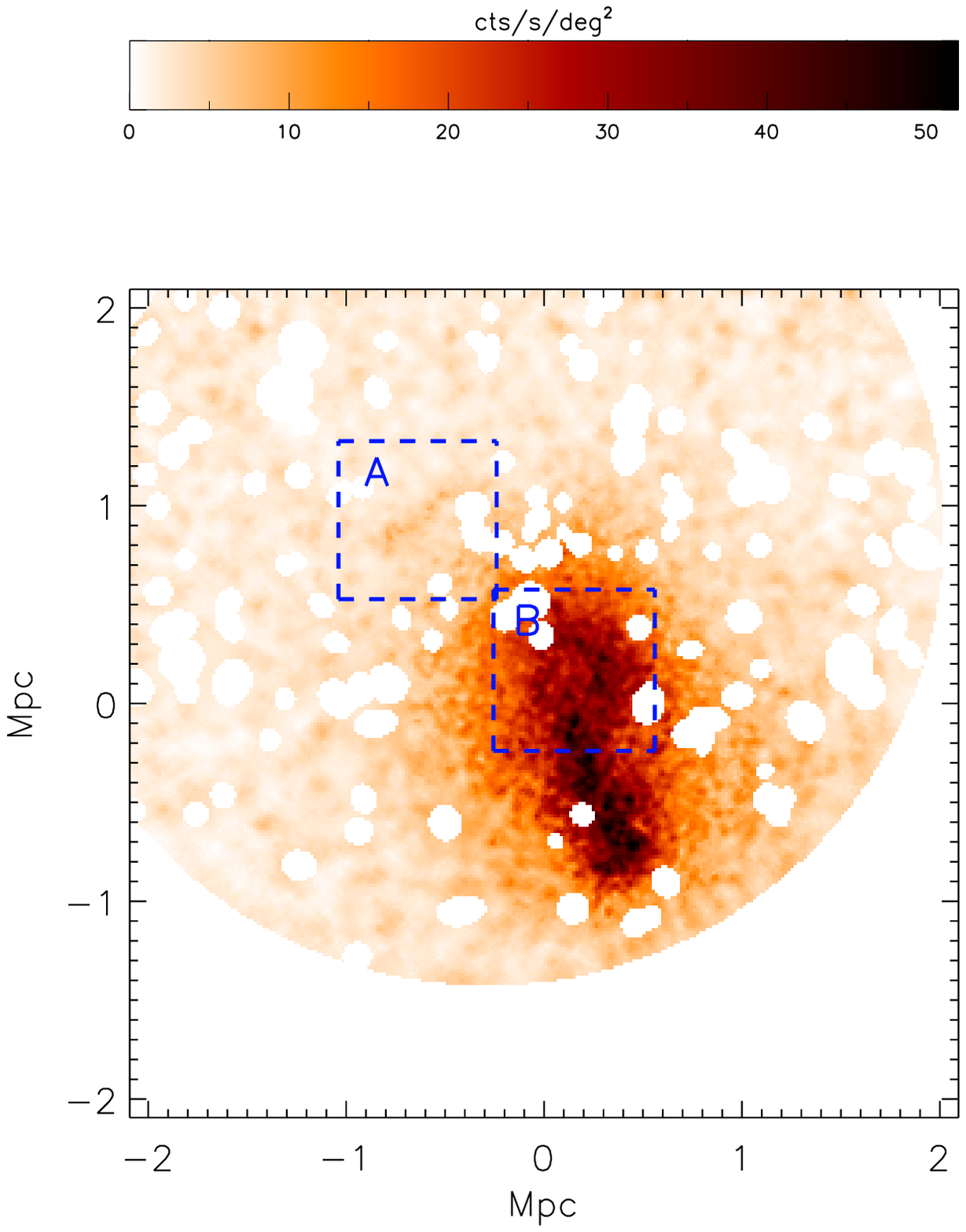}
  \vspace{-1cm}{
  \includegraphics[width=1.0\columnwidth,keepaspectratio=true,clip=true, trim=0.5cm 0cm 0.5cm 1.0cm]{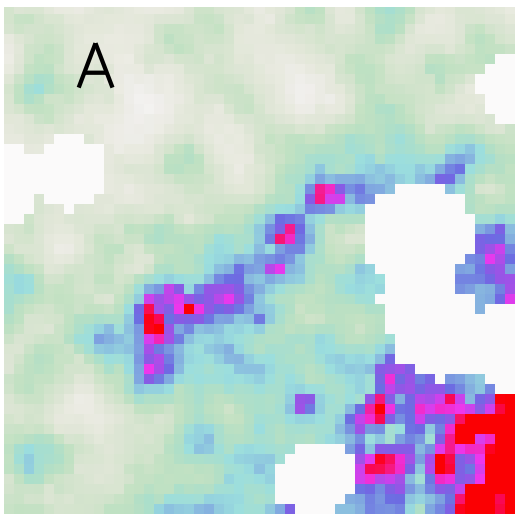}
  \includegraphics[width=1.0\columnwidth,keepaspectratio=true,clip=true, trim=0.5cm 0cm 0.5cm 1.0cm]{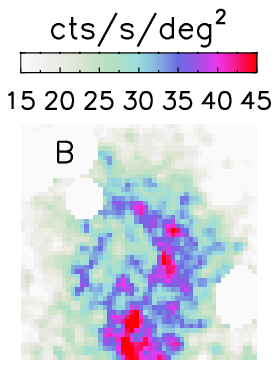}}
 \end{center}
 \caption{\emph{Top:} Exposure-corrected and instrumental background-subtracted \emph{XMM-Newton} EPIC surface brightness map of the cluster, in the energy band $0.5-4$ keV. The image was binned by 2, and adaptively smoothed with a minimum of 100 counts per bin. \emph{Bottom:} (A) Zoom-in on the ``smudge'' feature. (B) Zoom-in on the arc. The bottom two images have a side length of 0.8~Mpc.}
 \label{fig:xmmimg}
\end{figure*}

\section{Data analysis}
\label{s:analysis}

% Figure \ref{fig:rosat} presents the 1.4 GHz WSRT radio image, overlaid with \emph{ROSAT} X-ray contours. In the radio, a narrow, 2-Mpc-long relic is clearly visible to the north, while a smaller, less regular counter relic appears to the south. {\bf Numerical simulations suggest that the relics were created following a merger between two clusters with mass ratio 2:1, that collided in the plane of the sky with an impact parameter below $400$ kpc \citep{vanWeeren2011b}. Indeed, the elongated, unrelaxed X-ray morphology suggests that CIZA J2242.8+5301 is a merging galaxy cluster.} Hence, signatures of the merging process {\bf and indications of the merger geometry} should be observable in the ICM; possible examples include surface brightness edges, and complex temperature and pressure distributions.

\subsection{Imaging analysis}
\label{s:imanal}

To search for surface brightness edges, we created an image in the energy band $0.5-4$ keV, excluding the energy range $1.2-1.9$ keV, which contains emission from the instrumental Al K$\alpha$ and Si K$\alpha$ lines (see Section \ref{s:bkgmod} for details). At higher energies, the emission is dominated by background, hence our upper-energy cutoff.

Individual detector images were created using the tasks \texttt{mos-spectra} and \texttt{pn-spectra}, and combined with the \texttt{comb} routine. We used the {\sc xmm-esas} task \texttt{adapt\_900} to create an instrumental-background-subtracted, vignetting-corrected EPIC image, binned by 2, and adaptively smoothed with a minimum of 100 counts per bin. Due to the size of the \xmm\ FOV, we also limited the image to a circle of radius 12 arcmin around each detector's centre. The resulting image is shown in Figures \ref{fig:xmm-radio} and \ref{fig:xmmimg}. The X-ray morphology is strongly elongated along the north-south direction, evidence that the cluster is undergoing a merger along this axis. At $\sim 170$ kpc from the centre there is an arc of X-ray emission. Furthermore, a ``smudge" extending on a length of about 500 kpc is aligned with the relic on its easternmost side. To the east, the tip of the smudge coincides with the position of an AGN, seen as a bright radio point-like source in Figure \ref{fig:xmm-radio} \citep{vanWeeren2010}. Interestingly however, the AGN jets are oriented perpendicular to the relic, i.e. also perpendicular to the smudge, so it is not clear whether the X-ray emission originates from the AGN, or is instead associated with a density and temperature enhancement at a shock tracing the radio relic. The arc and the ``smudge'' feature are shown in Figure \ref{fig:xmmimg}.

\begin{figure*}
 \begin{center}
    \includegraphics[width=\textwidth]{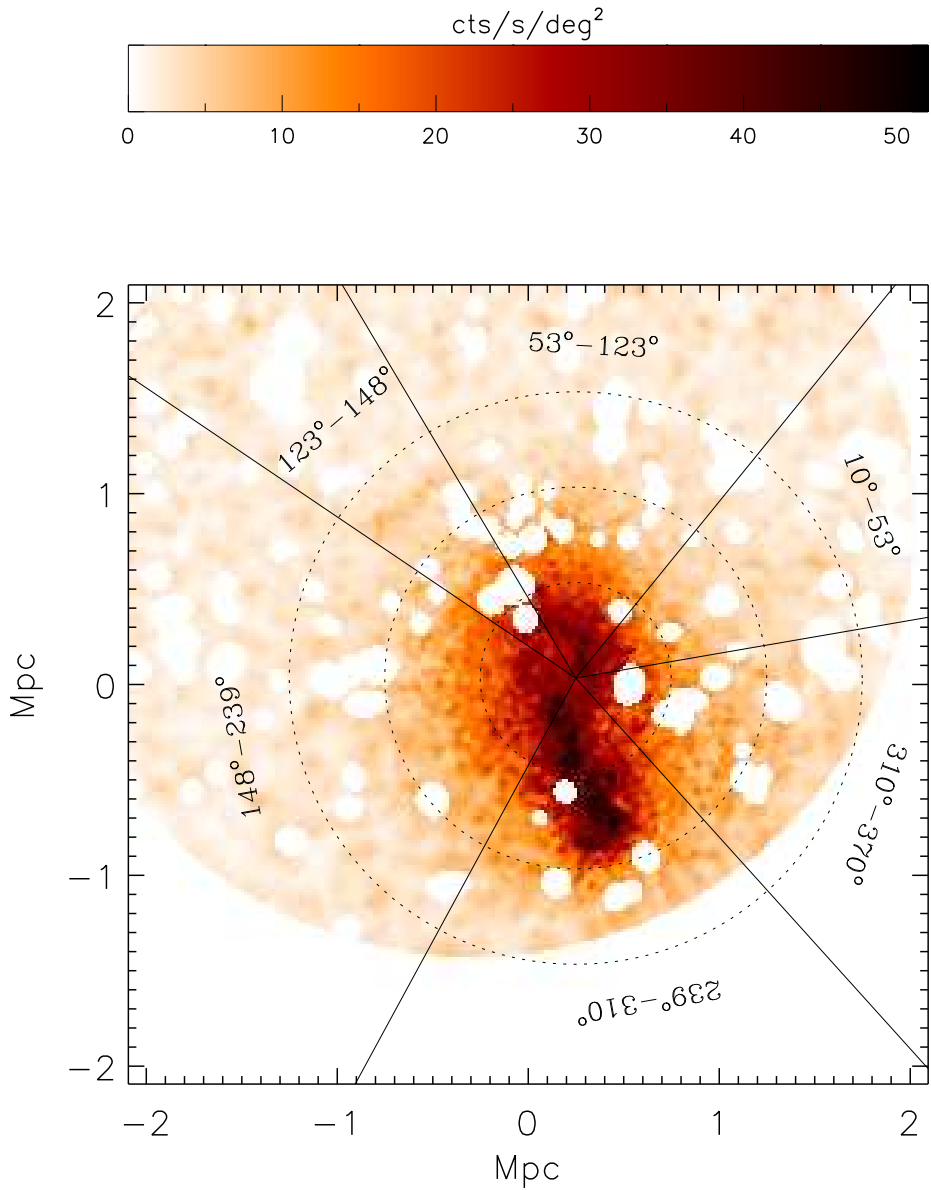}
 \end{center}
 \caption{Sectors used for creating surface brightness profiles. Dashed circles mark distances of 0.5, 1.0, and 1.5 Mpc from the centre, and are shown for easier comparison with the profiles in Figure \ref{fig:sx}.}
 \label{fig:annuli}
\end{figure*}

\begin{figure}
 \begin{center}
    \includegraphics[width=\columnwidth]{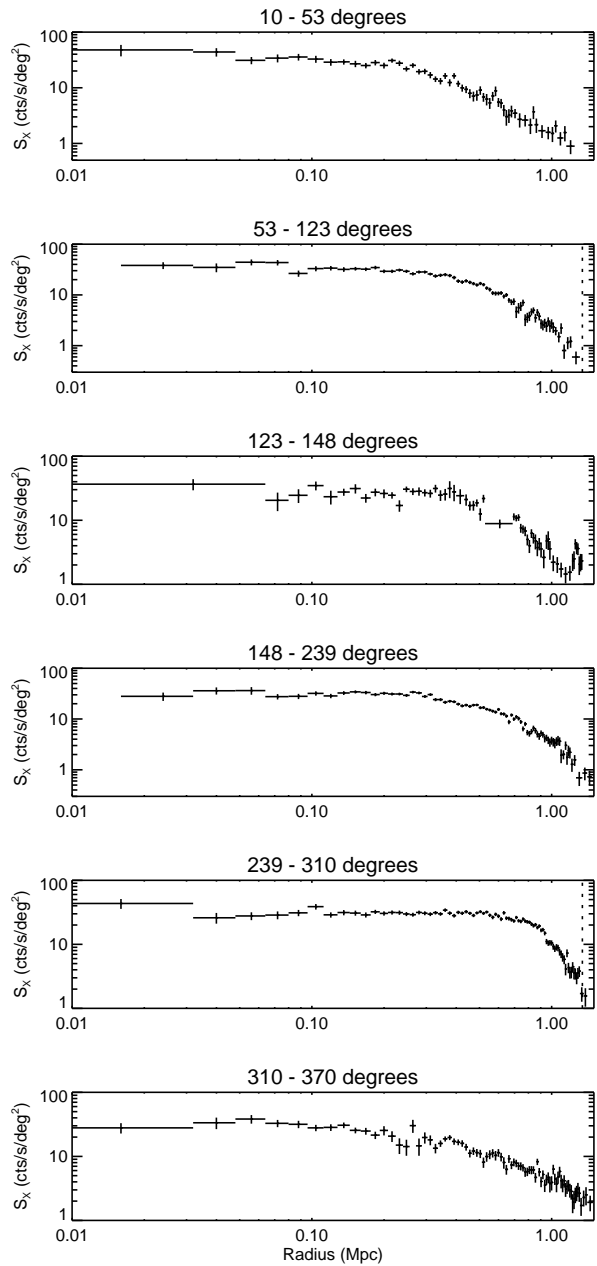}
 \end{center}
 \caption{Background-subtracted surface brightness profiles in the sectors shown in Figure \ref{fig:annuli}. Only bins with a ${\rm SNR} \ge 3$ are shown. Error bars are Poisson errors propagated in quadrature. Dotted lines show the positions of the relics based on the \emph{WSRT} 1.4 GHz observations. Profiles extend up to 1.5 Mpc.}
 \label{fig:sx}
\end{figure}

We examined the surface brightness of the cluster in six directions from the centre, as shown in Figure \ref{fig:annuli}, where the centre was defined as the geometrical centre of the double-relic system. For this, we created an instrumental-background-subtracted, vignetting-corrected EPIC surface brightness map in the energy band $0.5-4$ keV, binned by 2. To avoid large vignetting-correction uncertainties near the edges of the detectors, we limited the profiles to the inner 24 arcmin (a circle centred on each detector centre, and having a radius of 12 arcmin). We chose the northern pie sector in such a manner as to cover the northern relic, but avoid the northeastern AGN and the observed X-ray ``smudge'' (see Figure \ref{fig:xmmimg}, and also Section \ref{s:smudge}). East of the northern sector, we selected a narrower one which covers the ``smudge'' region. The other four sectors' angular widths were chosen such that surface brightness contours within each of them are roughly circular far from the cluster centre. We assumed a constant sky background surface brightness, and calculated its contribution from the crescent-moon shaped background region shown in Figure \ref{fig:rass}. The background surface brightness was then subtracted from the map. In each sector, the average surface brightness was calculated in partial annuli of constant signal-to-noise ratio (SNR), ${\rm SNR}=5$. All errors were propagated in quadrature. The background-subtracted surface brightness profiles are shown in Figure \ref{fig:sx}. We note that close to the centre, the complex substructure is not well-described by means of surface brightness profiles. However, our aim is to search for signatures of shock compression in the outskirts, where there is little substructure, and the merger's ellipticity can be safely ignored inside the selected sectors.

Beyond the northern relic, there is a single bin of ${\rm SNR} \approx 1.8$ (not shown in Figure \ref{fig:sx}) and surface brightness $0.08\pm 0.05$, so a factor of $\sim 7$ lower than the surface brightness in the (putative) post-shock region. Unfortunately, the ICM signal beyond the relic is too low to allow us to model a putative shock with a double power-law density profile, as it is customarily done \citep[e.g.][]{Macario2011}.  

Near the southern relic, we observe a surface brightness jump by a factor of $\sim 2-3$. Here, the small number of pre-shock bins with good SNR make modelling the density beyond the relic difficult. Instead, we model the pre-shock density assuming a beta-model described by the average $\beta$ power index derived by \citet{Eckert2012} for a sample of 31 galaxy cluster observed with \emph{ROSAT}. The method and results are presented in detail in Section \ref{s:shocks}.

In the narrow bin emcompassing the ``smudge'', there is a clear peak in the surface brightness at approximately 1.3~Mpc from the centre, coinciding with the position of the observed ``smudge''. The origin of this emission is discussed in Section \ref{s:smudge}.

\subsection{Background and foreground modelling}
\label{s:bkgmod}

The low surface brightness in cluster outskirts makes it essential to model accurately the different background and foreground (BF) components. Typically, there are four BF components that need to be considered: a quiescent particle background (QPB), produced by the interaction of high-energy particles with the detectors and other satellite components; unresolved background sources; the Galactic Halo (GH); and the Local Hot Bubble (LHB). However, \ciza\ is located at a low Galactic latitude, $|b|\approx 5^{\circ}$, where additional foreground components can be present. For example, \citet{Masui2009} detected thermal foreground emission with $T\sim 0.8$ keV in the Galactic direction ($235^{\circ}$, $0^{\circ}$), which they attribute most likely to a population of unresolved dM stars, whose number falls quickly with increasing Galactic latitude. \citet{Simionescu2011} found a foreground component with $T\approx 0.6$~keV towards the Perseus cluster, at Galactic latitude $|b|\sim 13^{\circ}$.\footnote{\citet{Simionescu2011} calls this component the ``hot foreground'' (HF), which is also how we will refer to it in this work.} \citet{Eckert2011} compared Suzaku \citep{George2009} and \emph{ROSAT} observations of the galaxy cluster PKS 0745-191 ($|b|\sim 3^{\circ}$), and found discrepancies between the surface brightness profiles measured with the two satellites; they attribute these discrepancies to additional background components that were not included in the analysis of \citet{George2009}. We note that the PKS 0745-191 \emph{Suzaku} data has been re-analysed recently by \citet{Walker2012} using new background observations near the cluster, and it was found that the inclusion of an additional background component with $T\approx 0.5$~keV provides the best fit to the background and eliminates the discrepancy between different instruments.

\begin{figure}
 \begin{center}
  \includegraphics[width=\columnwidth]{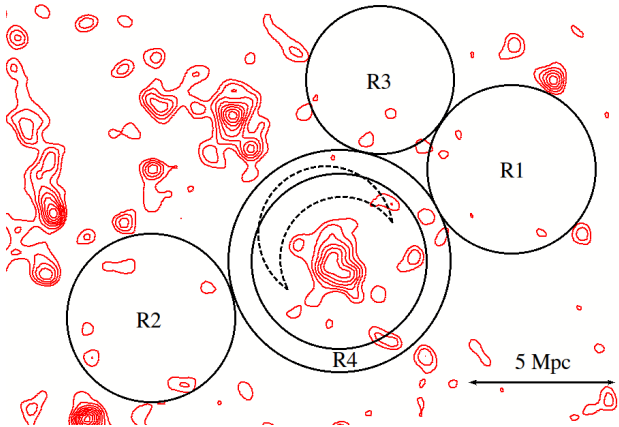}
 \end{center}
 \caption{\emph{ROSAT} PSPC contour map, showing equidistant contours. Overlaid in solid lines are the regions used for extracting \emph{ROSAT} All-Sky Survey (RASS) background spectra. The crescent-moon shaped region in dashed lines shows the region used for extracting the \xmm\ background spectra.}
 \label{fig:rass}
\end{figure}

To model the background near \ciza, as well as all the other spectra presented in this paper, we used version 12.7.1 of {\sc Xspec}. EPIC spectra were extracted from the background region shown in Figure \ref{fig:rass}, and grouped to a minimum of 30 counts/bin. To better constrain the soft components of the BF model, we also used four \emph{ROSAT} spectra extracted from an annulus around the cluster centre, and from three circles NW and SE of the cluster centre. We avoided the NE direction, where another galaxy cluster is visible in the \emph{ROSAT} image, and the S-SW directions, where the X-ray count rate is slightly higher. The positions and extent of the selected regions are summarized in Table \ref{tab:rass}, while Figure \ref{fig:rass} shows their location on the \emph{ROSAT} PSPC contour map.

{\sc xmm-esas} uses the CalDB files to model the QPB. However, it excises from the modelled QPB spectra the strong Al K$\alpha$ ($\approx 1.5$ keV; present in the MOS and pn spectra), Si K$\alpha$ ($\approx 1.75$ keV; present in the MOS spectra) and Cu ($\sim 8$ keV; present in the pn spectra) fluorescent instrumental lines. Therefore, these lines must either be added later to the spectral model describing cluster emission, or the energy sub-bands containing these lines need to be ignored in the fitting process. Fitting the Cu lines is difficult. Moreover, the pn data also has a strong low-energy tail at energies below $0.3-0.5$ keV. In consequence, we fitted the spectra in the energy band $0.5-7.0$ keV, excluding the $1.2-1.9$ keV MOS data points, and the $1.2-1.7$ keV pn data points.

Foreground emission was modelled with the sum of three APEC \citep{AtomDB} components, describing emission from collisionally-ionized diffuse gas: an unabsorbed component, corresponding to LHB emission, and two absorbed components, corresponding to GH and HF emission, respectively. The LHB temperature was fixed to 0.08 keV, which is the average of the temperatures found by \citet{Sidher1996} and \citet{KuntzSnowden2000}. We used the abundance table of \citet{angr1989}, and fixed the abundances of the foreground components to solar value. The redshifts were fixed to 0. The other APEC spectral parameters were free, but coupled between spectra. The cosmic X-ray background (CXB), consisting of emission from unresolved cosmological sources, was modelled with an absorbed power-law with a fixed photon index $\Gamma=1.41$ \citep{delucamolendi2004}. The CXB normalization was free, but coupled between the \xmm\ spectra. For the \rosat\ spectra, the CXB normalization was fixed to $8.85\times 10^{-7}$ photons\,keV$^{-1}$\,cm$^{-2}$\,s$^{-1}$\,arcmin$^{-2}$ \citep{Moretti2003}, as point sources were not excluded from these spectra. Photoelectric absorption was modelled with a PHABS component, and we used the photoelectric absorption cross-sections of \citet{Verner1996}.

\begin{table}
 \caption{Regions used for extracting \emph{ROSAT} background spectra. Region names correspond to the labels in Figure \ref{fig:rass}.}
\begin{center}
 \begin{tabular}{|l|c|c|c|}
  \hline
   Region & RA & DEC & Radius (deg) \\
  \hline
    R1 & $22^{\rm h}\,39^{\rm m}\,23^{\rm s}$ & $+53^{\circ}\,19'\,27''$ & $0.25$ \\
    R2 & $22^{\rm h}\,46^{\rm m}\,25^{\rm s}$ & $+52^{\circ}\,51'\,40''$ & $0.25$ \\
    R3 & $22^{\rm h}\,42^{\rm m}\,01^{\rm s}$ & $+53^{\circ}\,34'\,53''$ & $0.22$ \\
    R4 & $22^{\rm h}\,42^{\rm m}\,45^{\rm s}$ & $+53^{\circ}\,02'\,40''$ & $0.26$, $0.33$ \\
  \hline
 \end{tabular}
\end{center}
 \label{tab:rass}
\end{table}

\subsection{Temperature and pressure distribution}
\label{s:temperaturedistribution}

\begin{table}
 \caption{Best-fit background and foreground parameters.}
\begin{center}
 \begin{tabular}{lccc}
  \hline
   Component & $\Gamma$ & $T_{\rm X}$ & $N_{\rm X}$ \\
  \hline
    LHB & -- & $0.08^\dagger$ & $(3.4\pm 0.29) \times 10^{-7}$  \\
    GH  & -- & $0.14_{-0.0024}^{+0.0027}$ & $1.6_{-0.094}^{+0.095} \times 10^{-5}$ \\
    HF  & -- & $0.62_{-0.022}^{+0.023}$ & $(8.7\pm 0.59) \times 10^{-7}$ \\
    CXB$^{\dagger\dagger}$ & $1.41^\dagger$ & -- & $1.0_{-0.045}^{+0.046} \times 10^{-6}$ \\
  \hline
 \end{tabular}
\end{center}
 \hspace{-0cm}$\dagger$ {\footnotesize frozen}\\
 \hspace{-0cm}$\dagger\dagger$ {\footnotesize for comparison, \citet{delucamolendi2004} obtain a total CXB normalization (i.e., including both resolved and unresolved sources) of $(9.8\pm 0.35) \times 10^{-7}$~photons\,keV$^{-1}$\,cm$^{-2}$\,s$^{-1}$\,arcmin$^{-2}$ at 1~keV}
 \label{tab:bkgmod-freenh}
\end{table}

\begin{figure*}
 \begin{center}
  \includegraphics[width=0.49\textwidth,keepaspectratio=true,clip=true,trim=0.2cm 0cm 1.3cm 0cm]{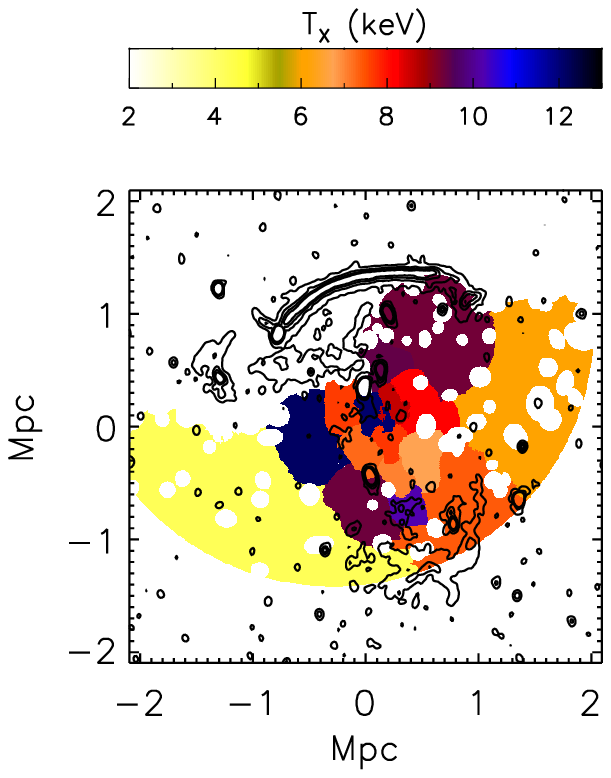}
  \includegraphics[width=0.49\textwidth,keepaspectratio=true,clip=true,trim=0.2cm 0cm 1.3cm 0cm]{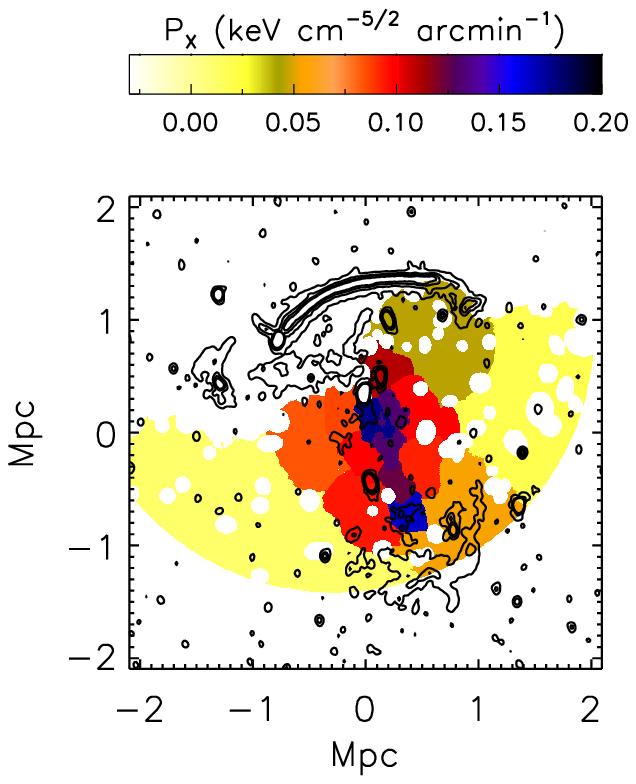}
 \end{center}
 \caption{Temperature and pseudo-pressure maps of the cluster. Spectra were extracted from bins that follow the surface brightness contours, and have a total of $\sim 3600$ counts after the subtraction of the instrumental background. The pressure was calculated as the product between the temperature and the square root of the normalization of the spectral component describing ICM emission. Overlaid are the radio contours also shown in Figure \ref{fig:xmm-radio}.}
 \label{fig:txpmap-freenh}
\end{figure*}

We used the contour binning algorithm developed by \citet{Sanders2006}\footnote{http://www-xray.ast.cam.ac.uk/papers/contbin/} to extract spectra with a total of $\sim 3600$ ICM plus BF counts. The code bins the data following the surface brightness contours of the map shown in Figure \ref{fig:xmmimg}. This approach is better suited than, for example, extracting spectra from annuli or partial annuli around the centre, given that the cluster has an extremely disturbed morphology, and that we are searching for density/temperature discontinuities in the ICM. We used a smoothing signal-to-noise of 30, and a geometric constraint value of 1.25 (see \citet{Sanders2006} for details). The small constraint value ensures that the bins do not become narrower than the \xmm\ PSF.

All spectra, including those of the background, were fitted simultaneously, in order to maximize the number of constraints and correctly propagate the statistical errors. The \xmm\ spectra were normalized by their respective geometrical areas (in units of sq arcmin) when defining the spectral model. The \rosat\ spectra are already in units of per sq arcmin. It was assumed that the ICM temperature in each bin is well-approximated by a single-temperature spectrum, which we modelled using an absorbed APEC component added to the background model described in the previous section. The ICM normalization was fixed to zero for the background spectra.

In the direction of the cluster (${\rm RA} = 22^{\rm h}\,42^{\rm m}\,53^{\rm s}$, ${\rm DEC} = +53^{\circ}\,01'\,05''$), in a circle of radius 1 degree, the weighted average atomic hydrogen column density ($N_{\rm H}$) listed in the Leiden/Argentine/Bonn (LAB) Survey of Galactic H{\sc i} \citep{Kalberla2005} is $3.22\times 10^{21}$ cm$^{-2}$. However, there is a large scatter in the LAB $N_{\rm H}$ values, with Galactic hydrogen column densities taking values in the range $(2.53-3.65)\times 10^{21}$ cm$^{-2}$. Therefore, the X-ray column density for each ICM spectrum was a free parameter in the fit. For the background spectra, there is some degeneracy between the column density and the foreground components, so the X-ray column density was fixed to the weighted average LAB value. We discuss in Section \ref{s:syserrors} how this affects our best-fit ICM parameters. The fit to the spectra had a $\chi^2/{\rm d.o.f.}$ of $4909.13/4358$. A spectral model without HF emission provided a poorer fit to the data, with $\chi^2/{\rm d.o.f.} = 5104.75/4360$. We also ran an F-test to check that the HF component is a reasonable addition to the model, and obtained a F-test probability $\sim 10^{-37}$. Therefore, a spectral model that includes HF emission provides the best fit to the data, as it was already discussed in more detail in Ogrean et al. (2012).

\begin{figure}
 \begin{center}
  \includegraphics[width=0.33\textwidth,angle=270]{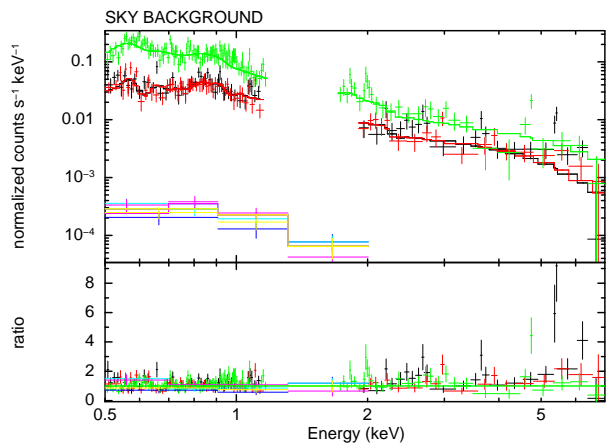}
 \end{center}
 \caption{Fit to the EPIC background spectrum extracted from the region marked in Figure \ref{fig:rass}. MOS1 and MOS2 data are plotted in black and red, respectively, while pn is plotted in green. The bottom pannel shows the ratio of the data to the model.}
 \label{fig:bkgspectrum-freenh}
\end{figure}

Figure \ref{fig:txpmap-freenh} shows the resulting temperature and pseudo-pressure maps. The pseudo-pressure was calculated as the product between the temperature and the square root of the normalization. The fit to the background spectrum is shown in Figure \ref{fig:bkgspectrum-freenh}, and in Table \ref{tab:bkgmod-freenh} we list the best-fit background parameters. Table \ref{tab:spectralfits} of the Appendix gives the temperature and normalization in each bin, together with the corresponding $1\sigma$ errors. Figure \ref{fig:binnum} in the Appendix shows the bin map, while Figure \ref{fig:binsp-freenh} presents the fitted spectra of all the bins. For comparison, we also fitted the spectra using an X-ray column density fixed to the average LAB value. The results of this fit are listed in Tables \ref{tab:bkgmod} and \ref{tab:spectralfits}, the corresponding spectra are shown in Figure \ref{fig:binsp}, while Figure \ref{fig:txpmap} shows the temperature and pseudo-pressure maps.

\section{Systematic uncertainties}
\label{s:syserrors}

\subsection{$N_{\rm H}$ uncertainties and sky background systematic errors}

\begin{figure*}
 \begin{center}
  \includegraphics[width=\textwidth,keepaspectratio=true]{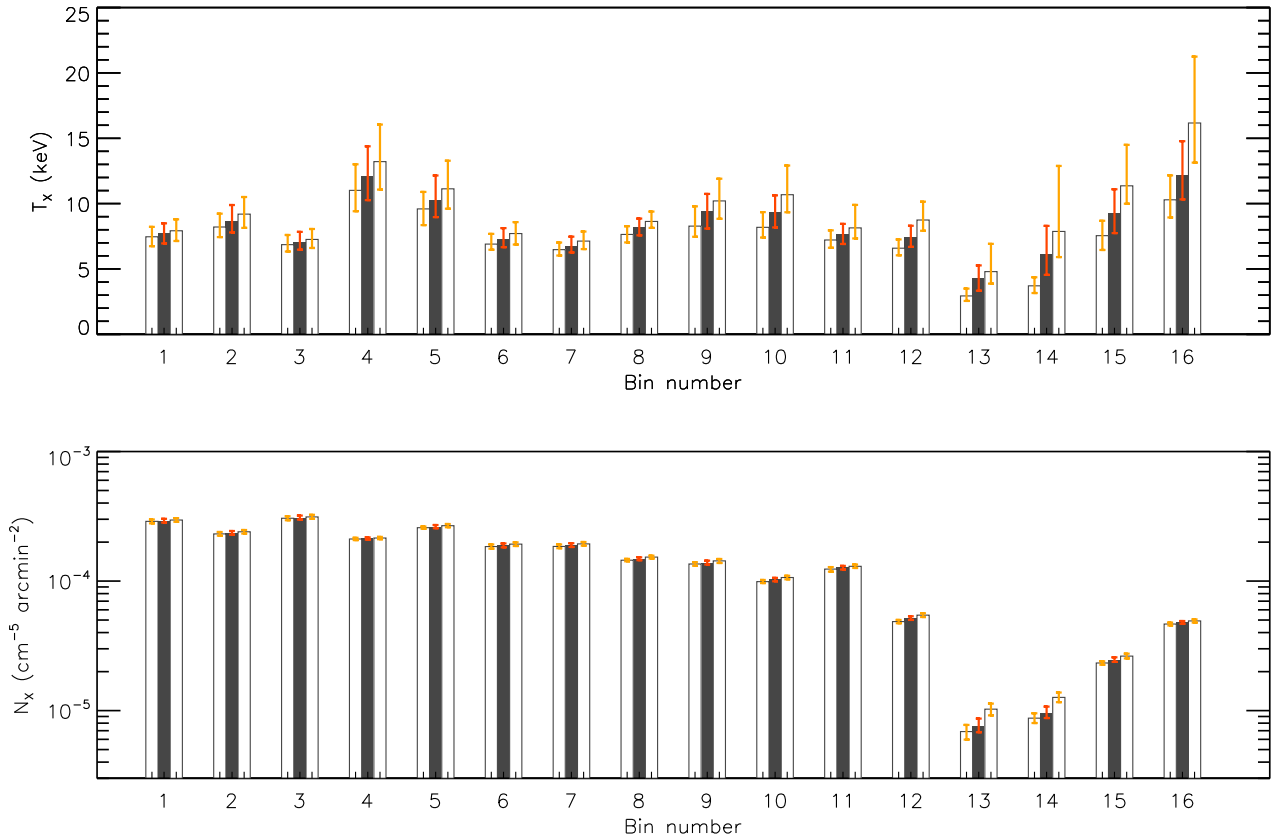}
 \end{center}
 \caption{The effect of sky background systematic errors and $N_{\rm H}$ uncertainties on the best-fit temperatures and normalizations. The minimum and maximum parameters for each bin were found by varying the sky background spectrum parameters within their {\sc min}$-1\sigma$ and {\sc max}$+1\sigma$ ranges, where the {\sc min}$-1\sigma$ and {\sc max}$+1\sigma$ values were the best-fit parameters and corresponding $\pm 1\sigma$ errors obtained, respectively, for the minimum and maximum $N_{\rm H}$ listed in the LAB Survey. See the text for details.}
 \label{fig:syserr}
\end{figure*}

Throughout the paper, the errors quoted are $1\sigma$ statistical errors. Therefore, the effects of systematic uncertainties on the fitted ICM spectra have so far been ignored. Furthermore, in fitting the spectra, we assumed the X-ray column density at the position of all five background regions was equal to the weighted average H{\sc i} column density listed in the LAB Survey, $N_{\rm H} = 3.22\times 10^{21}$ cm$^{-2}$, calculated in a circle of radius 1 degree centred on the cluster coordinates in the NASA/IPAC Extragalactic Database (NED), ${\rm RA} = 22^{\rm h}\,42^{\rm m}\,53^{\rm s}$, ${\rm DEC} = +53^{\circ}\,01'\,05''$. Because of the cluster's low Galactic latitude, there is a scatter of $20-25\%$ in the Galactic H{\sc i} column density within this $1^{\circ}$-radius cone. Local variations in $N_{\rm H}$ could affect our spectral measurements. Here, we examine the effect of systematic uncertainties and $N_{\rm H}$ fluctuations on the temperature and pressure distribution. To do this, we refitted the background spectra alone, with the X-ray column density fixed to the minimum and maximum Galactic hydrogen column densities listed in the LAB Survey within 1 degree of the cluster's centre, i.e. $N_{\rm H,\, min} = 2.53\times 10^{21}$ cm$^{-2}$ and $N_{\rm H,\, max}=3.65\times 10^{21}$ cm$^{-2}$. Using the {\sc min}$-1\sigma$ and {\sc max}$+1\sigma$ value of each 6 BF parameters, we then constructed 12 background models, each having one BF parameter fixed to its {\sc min}$-1\sigma$ or {\sc max}$+1\sigma$ value, while the other parameters were frozen to the best-fit values in Table \ref{tab:bkgmod-freenh}. Finally, the 16 ICM spectra were refitted using each of the 12 background models and a free X-ray column density. Figure \ref{fig:syserr} presents a comparison between the maximum and minimum best-fit temperatures and normalizations, and the best-fit values listed in Table \ref{tab:spectralfits}. Within the $1\sigma$ statistical errors of each measurement, the results are consistent for all bins, with the exception of bins 13 and 14. However, the large variations introduced by $N_{\rm H}$ uncertainties on the best-fit parameters of bins 13 and 14 does not change any of the main results presented in this paper. 

In conclusion, our main spectral results are robust, in spite of uncertainties on the X-ray column density in the direction of the background regions. Even by varying the $N_{\rm H}$ corresponding to the background regions by more than $1\times 10^{21}$ cm$^{-2}$, the temperature and normalization distributions are qualitatively similar to those shown in Figure \ref{fig:txpmap-freenh}.

\subsection{Uncertainties on the QPB}

\begin{figure*}
 \begin{center}
  \includegraphics[width=\textwidth,keepaspectratio=true]{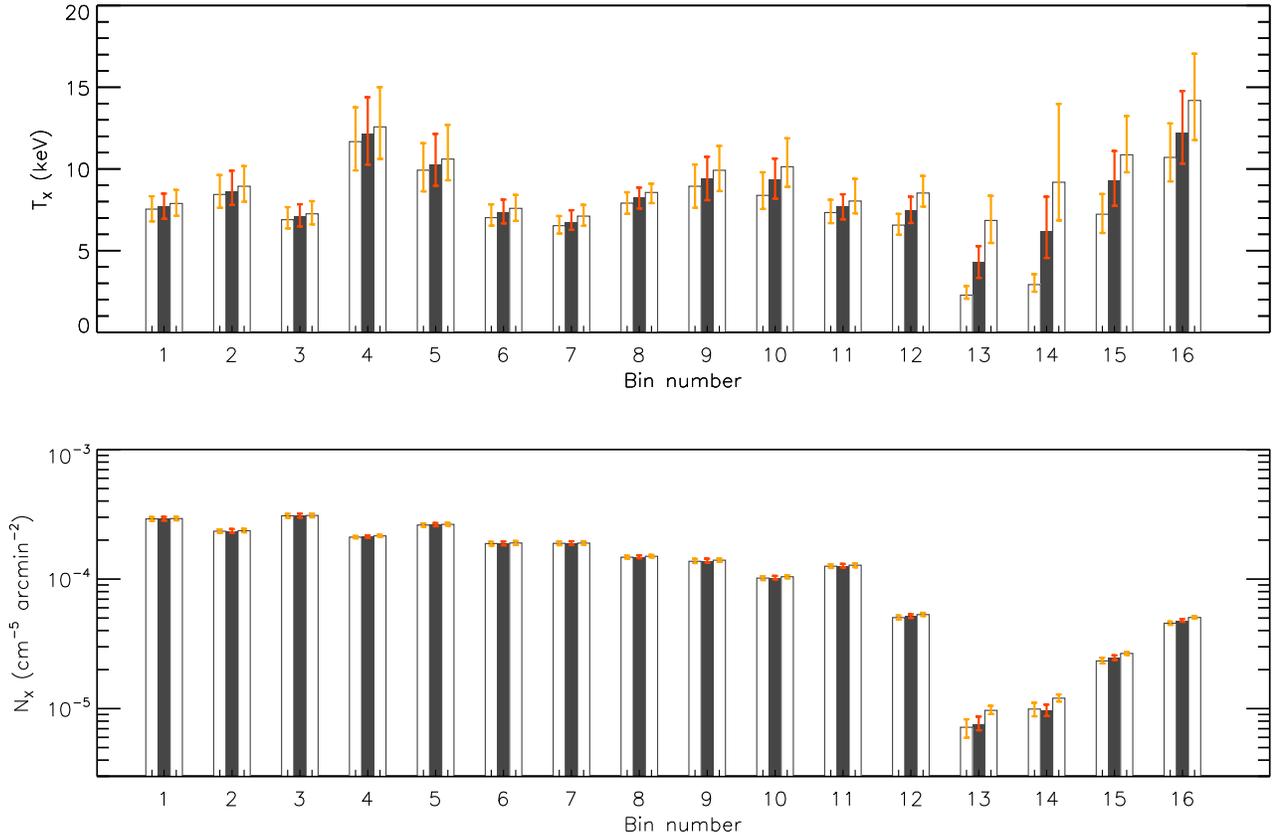}
 \end{center}
 \caption{The effect of a $5\%$ instrumental background uncertainty on the best-fit temperatures and normalizations.}
 \label{fig:qpberr}
\end{figure*}

We also examined the spectral effects of a $5\%$ uncertainty on the normalization of the quiescent particle background. We first varied the BACKSCAL keyword of the instrumental background spectra associated with the sky background region by $\pm 5\%$, and derived two new foreground+CXB best-fit emission models. The best-fit parameters of these fits were then used to refit each ICM spectrum, while at the same time changing the normalization of the corresponding QPB spectra associated with the ICM bins by $\pm 5\%$. MOS and pn QPB normalizations were raised and lowered simultaneously. Figure \ref{fig:qpberr} shows the changes on the best-fit parameters of each ICM bin, by comparing the new best-fit values with those summarized in Table \ref{tab:spectralfits}. Again, the best-fit parameters of bins 13 and 14 depend strongly on background uncertainties, and a $5\%$ error in the normalization of the quiescent particle background yields temperatures inconsistent with the best-fit results presented in Section \ref{s:temperaturedistribution}. However, the best-fit parameters of the other bins are consistent with our previous results.

\subsection{Solar wind charge exchange}

\begin{figure}
 \begin{center}
  \includegraphics[width=\columnwidth,keepaspectratio=true]{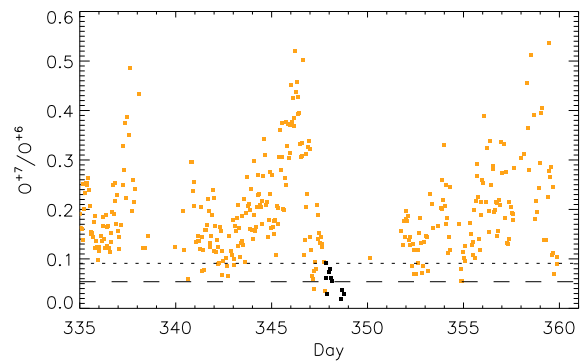}
 \end{center}
 \caption{O$^{+7}$/O$^{+6}$ ratio. Black points correspond to measurements taken during the time of the \xmm\ observation of CIZA~J2242.8+5301, while the dashed and dotted lines show, respectively, the mean and maximum of these data points.}
 \label{fig:swcx}
\end{figure}

Solar wind charge exchange (SWCX) photons, released in the aftermath of collisions between atoms and solar wind ions, sometimes affect X-ray observations below $\sim 2$~keV. The main lines appearing in the SWCX spectrum are C{\sc vi} (0.37, 0.46~keV), N{\sc vi} (0.42~keV), O{\sc vii} (0.56~keV), O{\sc viii} (0.65, 0.80~keV), Ne{\sc ix} (0.92, 1.15~keV), and Mg{\sc xi} (1.34~keV). Of these, the carbon and oxygen lines are generally the strongest \citep[e.g.][]{Snowden2004,CarterSembay2008,Carter2011}. \citet{Snowden2004} has shown that periods of enhanced SWCX are associated with an increase in the O$^{+7}$/O$^{+6}$ ratio. In Figure \ref{fig:swcx}, we show the hourly O$^{+7}$/O$^{+6}$ ratio over a period of 27 days (12 days prior and following the \xmm\ observation). The solar wind data was obtained with the SWICS (Solar Wind Ion Composition Spectrometer) instrument on board of the \emph{ACE} (Advanced Composition Exporer) satellite. We only show FLAG=0 (i.e., good quality) data points. As Figure \ref{fig:swcx} shows, during the \xmm\ observation the O$^{+7}$/O$^{+6}$ ratio was among the lowest recorded over almost a whole month. Therefore, SWCX emission is negligible.

\section{Evidence for shocks in the ICM}
\label{s:shocks}

In Figure \ref{fig:sx}, a surface brightness discontinuity can be seen to the south, at approximately 1.3~Mpc from the centre of the sectors in Figure \ref{fig:annuli}. We modelled it assuming spherical symmetry, and an underlying density profile described by two power-laws with different normalizations and indices, one for each side of the discontinuity:
\begin{eqnarray}
  n_2(r) & = & n_{\rm shock} \left(\frac{r}{r_{\rm shock}}\right)^{-\gamma_1} \,\, , \, r \le r_{\rm shock} \nonumber \\
  n_1(r) & = & \frac{1}{C}\,n_{\rm shock} \left(\frac{r}{r_{\rm shock}}\right)^{-\gamma_2} \,\, , \, r > r_{\rm shock}
\end{eqnarray}
where $C$ is the shock compression, $r_{\rm shock}$ is the cluster-centric distance at which the density discontinuity is observed, and $n_{\rm shock}$ is the electron number density of the post-shock gas. Starting from this density model, the surface brightness, $S_{\rm X}$, was calculated by projecting the emissivity along the line of sight, $S_{\rm X} = \int_{-L}^{L} \epsilon {\rm d}l$, with the emissivity, $\epsilon$, defined as:
\begin{eqnarray}
  \epsilon = n_{\rm e}n_{\rm i} \Lambda (T,Z)
\end{eqnarray}
where the ion density is $n_{\rm i}=n_{\rm e}/1.2$, and $\Lambda$ is the cooling function,
\begin{eqnarray}
\Lambda (T,Z) = C_1(kT)^{\alpha}+C_2(kT)^{\beta}+C_3 
\end{eqnarray}
in units of $10^{-22}\,\,\,{\rm erg\,cm^3\,s^{-1}}$, and with $\alpha = -1.7$, $\beta = 0.5$, $C_1 = 8.6\times 10^{-3}$ ${\rm erg\, cm^3\, s^{-1}\, keV^{1.7}}$, $C_2 = 5.8\times 10^{-2}$ ${\rm erg\, cm^3\, s^{-1}\, keV^{-0.5}}$, and $C_3 = 6.4\times 10^{-2}$ ${\rm erg\, cm^3\, s^{-1}}$ for $Z=0.3Z_{\sun}$ \citep[e.g.,][and references therein]{Ruszkowski2004}. $L$ is the maximum distance from the centre of the cluster, measured along the line of sight; since we assumed spherical symmetry, this distance is equal to the maximum projected distance to which the surface brightness profile is calculated. Because the number of points in the putative pre-shock region of the surface brightness profile is small, the power-law index of the density model on this side of the discontinuity was fixed to 2.466, which corresponds to three times the average $\beta$ index obtained by \citet{Eckert2012} when fitting a beta model to the profiles of 17 non-cool core (NCC) \rosat\ clusters between $0.65$ and $1.3r_{\rm 200}$. Furthermore, the distance of the jump from the centre was fixed to the average of the two radii on both sides of the observed discontinuity. In conclusion, the fits had three free parameters: $C$, $n_{\rm shock}$, and $\gamma_2$.

\begin{figure}
 \begin{center}
  \includegraphics[width=0.33\textwidth,angle=270]{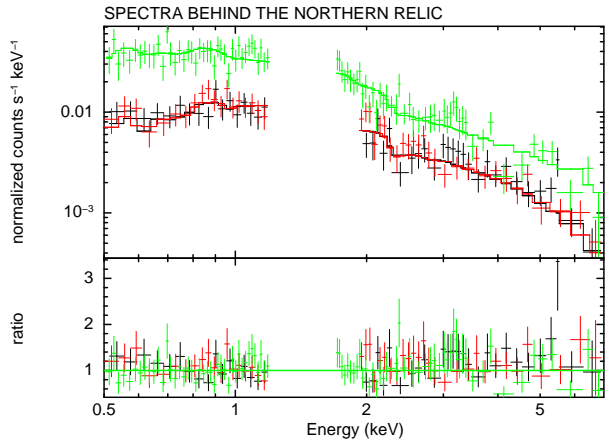}
 \end{center}
 \caption{EPIC spectra extracted from the (putative) post-shock region behind the northern relic, and the best-fit thermal model. MOS1 and MOS2 data are plotted in black and red, respectively, while pn is plotted in green. The bottom pannel shows the ratio of the data to the model.}
 \label{fig:postshock}
\end{figure}

The southern surface brightness profile crosses the radio relic. However, assuming a shock is present at the relic, the expected pre-shock region (i.e., the region right in front of the relic) lies beyond 12 arcmin from the detectors' centre, and it is not included in our imaging analysis. Instead, the observed surface brightness discontinuity at $\sim 1$ Mpc is due to the fainter region SE of the relic, at angles roughly between $240^{\circ}$ and $270^{\circ}$. We assumed a temperature of 7 keV on both sides of the discontinuity, which is an acceptable estimate based on the temperature map in Figure \ref{fig:txpmap-freenh}. The best-fit to the surface brightness profile has $\chi^2/{\rm d.o.f.} = 13.9/19$, and yields $C = 1.35\pm 0.14$, corresponding to a Mach number of $1.24\pm 0.097$. We tested the effect of varying $\gamma_1$ on the resulting Mach number. The highest $\beta$ derived by \citet{Eckert2012} is $\beta=0.963\pm 0.054$, for cool core (CC) clusters whose profiles were fitted with a beta model in the radial range $0.65-1.3r_{\rm 200}$, while the lowest is $\beta = 0.677\pm 0.002$, for NCC cluster profiles fitted in the radial range $0-0.7r_{\rm 200}$. Therefore, we varied $\gamma_1$ between 2.025 and 3.051, and found Mach numbers in the range of $1.2-1.3$. If a Mach number of $\sim 1.3$ reflects the strength of the shock at the southern relic, it would be the lowest Mach number shock known at a relic. This would be particularly surprising so far from the merger centre, where the sound speed is expected to be low, especially with the southern relic having a counter-relic that is expected to trace one of the strongest merger shocks ($\mathcal{M}\sim 4-5$ based on the radio-derived Mach number for the northern relic, \citet{vanWeeren2010}). 

It is also interesting that the shock extends, to the east, beyond the southern relic. A similar feature has been previously observed by \citet{Mazzotta2011} in RX\,J1314.4-2515, in which a shock extends beyond the western relic to the north and south. It is possible that the shock strength varies across the shock boundary, and the Mach number of the putative shock across the southern relic is actually higher than 1.3. However, studying this would require further observations, this time focused on the southern relic.

In the north, the temperature peaks at $\approx 9$~keV near the western tip of the relic, as can be observed in Figure \ref{fig:txpmap-freenh}. However, none of the bins provides an adequate representation of the (putative) post-shock gas. Therefore, we measured the temperature in a partial annulus extracted from the $53^{\circ}-123^{\circ}$ sector shown in Figure \ref{fig:annuli}, between radii of 0.75 and 1.35 Mpc; we note that this region overlaps most of the relic. The spectral model was the same as that described in Section \ref{s:temperaturedistribution}, but the background parameters were kept fixed to the values listed in Table \ref{tab:bkgmod-freenh}. The best-fit temperature was found to be $9.2_{-1.6}^{+2.7}$ keV, and the fit had $\chi^2/{\rm d.o.f.} = 225.43/244$. The fitted spectra are shown in Figure \ref{fig:postshock}. Therefore, we see the temperature slightly increasing from $7-8$~keV at the centre of the merger, to approximately 9~keV behind the northern relic. The radio spectral index at the front of the relic predicts, in the linear regime, a Mach number of $4.6_{-0.9}^{+1.3}$. If this Mach number matched the density and temperature jumps across the relic, then we would expect a shock compression of approximately 3.5, and a pre-shock temperature of $\sim 1$ keV. \citet{Akamatsu2011}, using \emph{Suzaku} observations of the cluster, measured a pre-shock temperature of approximately 1.6~keV. However, their post-shock temperature is only $6.7$ keV, much below the temperature measured behind the relic using our \xmm\ observations. As also pointed out in Ogrean et al. (2012), most of this difference is caused by an incorrect background model used by \citet{Akamatsu2011}, who omitted the HF component. At this point, it is unclear how an additional foreground component would affect the \emph{Suzaku} measurement of the pre-shock temperature.

The low count statistics of our data do not allow us to constrain the density or temperature beyond the northern relic. However, the temperature map indicates an increase in temperature at the position of the relic. Such a temperature profile towards merger shocks is not surprising. As the shock propagates from the centre towards the outskirts, the downstream gas furthest from the current shock position has had the longest time to cool. Assuming that the ICM can be approximated by an isothermal gas sphere, the ICM that was traversed most recently by the shock will have the highest temperature \citep{Roettiger1997}. However, these approximations are simplistic, and substructure, projection effects, as well as the underlying cluster temperature profile can erase the temperature gradient. In fact, a positive temperature gradient in the direction of the relic has not been observed in the two relic-hosting mergers with confirmed shocks. 

% We speculate that in CIZA J2242.8+5301, the merger geometry and a large shock Mach number result in projected gas temperatures that dominate over the underlying cluster temperature profile, which also might had been flattened by the mixing of the clusters' ICM.} 

In conclusion, we find evidence of a weak shock with $\mathcal{M}\sim 1.2-1.3$ near the southern relic and, interestingly, the shock extends beyond the SE tip of the relic. In the north, the temperature is highest at the relic, peaking at roughly 9~keV. Unfortunately, our \xmm\ observations do not allow us to constrain the density or the temperature in front of the northern relic, and therefore we are not able to measure the Mach number of the putative shock.

\begin{figure}
 \begin{center}
  \includegraphics[width=\columnwidth,keepaspectratio=true,clip=true,trim=0.5cm 0cm 1cm 0cm]{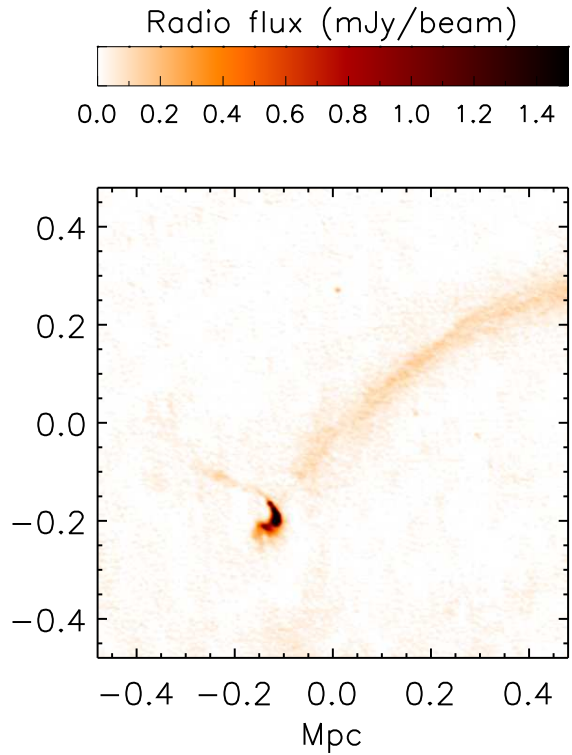}
 \end{center}
 \caption{\emph{GMRT} 1280 MHz radio flux map, focused on the AGN at the eastern tip of the northern relic in CIZA J2242.8+5301. While the AGN is positioned in continuation of the relic, it is detached from it, and the AGN jets are pointing towards NE, perpendicular on the relic.}
 \label{fig:agn}
\end{figure}

\begin{figure*}
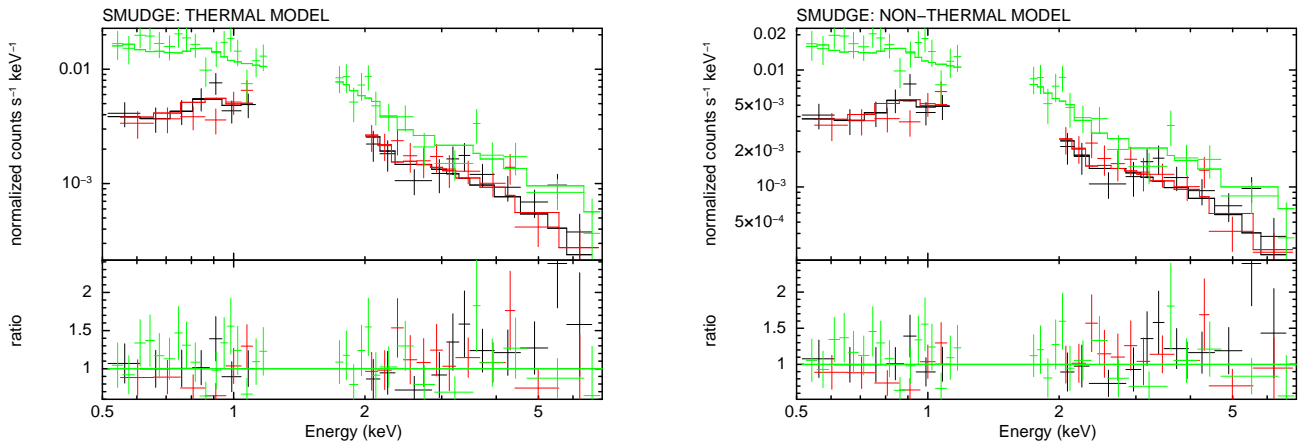

 \begin{center}
  \hspace{-1.0cm}\includegraphics[width=0.33\textwidth,angle=270]{smudge-th.ps}
  \hspace{0.8cm}\includegraphics[width=0.33\textwidth,angle=270]{smudge-pow.ps}
 \end{center}
 \caption{The ``smudge'' spectra, fitted with a thermal, and non-thermal model. MOS1 and MOS2 data are plotted in black and red, respectively, while pn is plotted in green. The bottom pannel shows the ratio of the data to the model.}
 \label{fig:agn-sp}
\end{figure*}

\begin{table}
 \caption{Best-fit parameters to the ``smudge'' spectra.}
\begin{center}
 \begin{tabular}{lcc}
  \hline
                   & \textsc{thermal model} & \textsc{non-thermal model} \\
  \hline
   $T_{\rm X}$ & $9.1_{-2.3}^{+7.7}$ & -- \\
   $\Gamma$ & -- & $1.6\pm 0.14$ \\
   $\mathcal{N}_{\rm X}$ & $(1.4\pm 0.12) \times 10^{-5}$ & $3.6_{-0.60}^{+0.68}\times 10^{-6}$ \\
  \hline
   $\chi^2/{\rm d.o.f.}$ & $86.88/96$ & $87.79/96$ \\
  \hline
 \end{tabular}
\end{center}
 \label{tab:agn}
\end{table}

\section{The northeastern AGN}
\label{s:smudge}

The \xmm\ surface brightness map shows a faint, half-Mpc-long ``smudge'' located northeast from the cluster. This coincides with an AGN identified by \citet{vanWeeren2010} in the radio. We extracted EPIC spectra in a partial annulus of area $\sim 6$ arcmin$^2$ overlapping the X-ray ``smudge'', and fitted them with two different spectral models: a non-thermal model described by a redshifted power-law, and a thermal model, described by an absorbed APEC model. The photon indices, temperatures, and normalizations were coupled between detectors, while the background was fixed to the best-fit background model summarized in Table \ref{tab:bkgmod-freenh}. The results of the fits are summarized in Table \ref{tab:agn}, and the fitted spectra are shown in Figure \ref{fig:agn-sp}. The errors on the thermal model are quite large and tip the balance toward non-thermal emission from the smudge. However, we cannot clearly differentiate between thermal and non-thermal emission. Unfortunately, the \xmm\ data does not allow us to fit a combined thermal plus non-thermal model to the ``smudge'' spectra.

It is not clear from the present observations whether the ``smudge'' is associated with the AGN or with the relic. In low frequency radio observations, the AGN appears detached from the relic, as can be seen in Figure \ref{fig:agn}. Its jets are oriented perpendicular to the ``smudge'', while the ``smudge'' is almost perfectly aligned with the southern boundary of the eastern tip of the northern relic. It is possible that a radio galaxy is ram-pressure stripped as it is moving through the ICM, possibly driven by the motions triggered by the merger, and this might be what we are observing in the X-ray. However, if this is the case, it is difficult to explain why at least one jet is so clearly oriented perpendicular to the apparent direction of motion. 

If the ``smudge'' was caused by a merger shock, we would be more likely to see it at the front of the relic, and it is unclear why the emission would be so enhanced only on one side of the shock. One possibility is the existence of a cosmic filament oriented roughly NE-SW. The density inside this filament would be higher than the density of its surroundings, which would increase bremsstrahlung emission in the part of the shock that crosses the filament. The shock speed along the filament would also be lower, which would curve the eastern tip of the relic inwards, and could diffuse the synchrotron-emitting particles; indeed, the fact that the geometrical centre of the relic system, assuming sphericity, is shifted slightly to the west, and that the radio structure S-SE of the relic is more filamentary could be an indication of this geometry. On the other hand, the magnetic field in the ICM region crossed by the filament would also be higher than in its surroundings. The synchrotron power scales as $P_{\rm syn} \propto B^2$, where $B$ is the magnetic field. The radio surface brightness of the northern relic in CIZA J2242.8+5301 is surprisingly uniform, therefore, assuming a filament crosses the eastern part of the relic, the lower shock speed inside the filament should compensate for the higher magnetic field.

\section{The merger scenario}
\label{s:mergerscenario}

Numerical simulations by \citet{vanWeeren2011b} predict that CIZA J2242.8+5301 is a merger occurring in the plane of the sky, between two almost equal-mass clusters (mass ratio $2:1$). This simple geometry minimizes complexities introduced by projection effects, and would indicate that the cluster is a textbook example for studying cluster mergers and large-scale shocks. However, also from numerical simulations, we know that such a merger scenario would result in two outward-moving shocks (advancing through the ICM perpendicular to the merger axis), two radio relics tracing these shocks, and a relatively symmetric temperature and pressure distribution. While the pressure map of CIZA J2242.8+5301, shown in Figure \ref{fig:txpmap-freenh}, is roughly the same on both sides of the merger axis, the temperature map is far from symmetric. Two hot bins enveloped in cooler plasma are seen north and south of the centre. The existence of these hot spots is indeed predicted by numerical simulations of cluster mergers, which show that the cores of the merging clusters can be heated to temperatures above the post-shock temperature, during the late phases of the merger event \citep[e.g.,][]{vanWeeren2011b}. However, to the east, the temperature increases from the centre into a hot ``wall'' with temperatures of $9-12$ keV. The existence of this hot region is puzzling, as the gas is not associated with a radio relic, and we do not identify a surface brightness discontinuity at its position. Furthermore, the presence of a hot eastern ``wall'' is in apparent contradiction with the simple merger geometry indicated by numerical simulations, suggesting a more complicated merging scenario, or a lack of understanding on our part of the complex structures formed during real cluster mergers.

One possibility is that the merger event had a small impact parameter. This scenario resembles that shown in fig. 2 ($t\approx 2.4$ Gyr) of \citet{vanWeeren2011b}, if a small counterclockwise rotation is applied to the image, such that the relics are aligned along the N-S direction. In this merger geometry, however, a ``wall'' of cold gas should be present east of the merger axis, roughly 0.5 Mpc from the merger centre (fig. 3, right, of \citet{vanWeeren2011b}). This is the opposite to what is observed in the temperature map, as the gas 0.5 Mpc east of the cluster centre is in fact very hot. If the ``wall'' of cold gas predicted by numerical simulations actually forms in real mergers, then in seems more likely that it is present west of the centre of CIZA J2242.8+5301, while to the east, the ICM is heated by a process or event that is unaccounted for in the numerical simulations (e.g., a triple cluster merger, with two clusters colliding in the plane of the sky, and a third one moving roughly along the line of sight). If this is the case, then the main merger event involved a collision between two almost equal mass clusters (mass ratio between 1:1 and 2:1), with the less massive cluster entering the gravitational well of the more massive cluster from the SE. There are indications that this scenario is also supported by the surface brightness distribution, which appears slighly broadened in the SW-NE direction, compared to the SE-NW. The constraints set by the X-ray morphology on the merger scenario will be discussed in more detail in Ogrean et al. (in prep.), using high resolution maps from a deep (200 ks) \emph{Chandra} observation.

{To summarize, although we are not able to measure the temperature beyond the spectacular northern relic, the temperature distribution to the north and south of the cluster centre is strikingly similar to that predicted by numerical simulations of merging clusters triggering large-scale shocks. However, while for CIZA J2242.8+5301 numerical simulations suggest a ``simple'' merger between two nearly-equal mass clusters colliding in the plane of the sky, this scenario is not supported by our X-ray results. We observe a very asymmetric temperature distribution, with a ``wall'' of hot plasma separating the merger centre from the eastern-most outskirt regions. The most likely merger scenario is that in which a less massive cluster entered the gravitational well of the more massive cluster from the SW, with a small impact parameter.}

\section{Summary}
\label{s:conclusions}

Based on radio data and numerical simulations, CIZA J2242.8+5301 appears to be a textbook example of a merging galaxy cluster hosting a double-relic system. The predicted merger geometry is simple: two clusters of almost equal masses merging along the north-south direction and essentially in the plane of the sky, which collided with a small impact parameter \citep{vanWeeren2011b}. However, this simple picture is only partially supported by the \xmm\ results presented above. Below we summarize our main findings:

\begin{itemize}
  \item In the X-ray, the cluster does indeed show an extremely disturbed, N-S-elongated morphology, indicative of a merger along the N-S direction. The southern part of the main X-ray structure is linear and brightest toward its southern tip, reminiscent of a bullet-like morphology. From the N-NE tip of this linear structure, the emission curves in a west-pointing arc of enhanced surface brightness. This X-ray morphology is more complex than predicted by numerical simulations.
  \item Because of the relatively high instrumental background level of \xmm, we were not able to characterize the surface brightness profile beyond the spectacular northern relic, and therefore find no evidence of shock compression near the relic. To the south, however, slightly east of the southern relic, we find evidence of a shock of Mach number $\sim 1.2-1.3$. Due to the size of the \xmm\ FOV, we do not have data immediately beyond the relic. Nevertheless, we speculate that a shock is present at the southern relic, and that what we observe is its extension beyond the detected boundaries of the relic. Possible reasons for different relic and shock lengths include Mach number fluctuations across the shock, coupled with weak cosmic ray injection efficiency at low Mach number shocks, or simply detector limitations.
  \item While numerical simulations of mergers with low mass ratios and low impact parameters predict relatively symmetric temperature distributions, this is not what is observed in CIZA J2242.8+5301. The temperature increases at the northern relic, as expected for shocks at which the increase in gas temperature dominates over the underlying temperature profile of the cluster. A smaller increase in temperature is also visible near the southern relic. What is more surprising is a ``wall'' of hot gas located east of the cluster centre. This ``wall'' is not associated with a radio relic. However, it appears to extend into the hot region behind the northern relic, forming approximately a semi-ellipse of hot plasma around the cluster. This temperature distribution is not what numerical simulations predict for a simple merger geometry.
  \item A ``smudge'' of enhanced X-ray emission can be seen aligned with the easter tip of the northern radio relic. Part of the smudge also overlaps the position of an AGN visible in the radio, whose jets point away from the cluster centre, perpendicular to the relic. It is not clear whether the ``smudge'' is associated with density compression at the relic, or with the AGN. Its spectrum appears to be better described by a non-thermal model. However, given the orientation of the AGN jets and the fact that it is clearly detached from the relic, we consider that the ``smudge'' is more likely a result of shock compression at the relic. The fact that it is visible only along $\sim 1/5$ of the length of the relic could be the result of cosmic structure (e.g., a cosmic filament), or a sign that the properties of the (putative) shock at the northern relic vary strongly across the relic's length.
\end{itemize}

\section*{Acknowledgments}

GAO thanks Franco Vazza, Annalisa Bonafede, Dominique Eckert, Silvano Molendi, and Hans B\"ohringer for helpful discussions, and also the \emph{XMM-Newton} Help Desk, in particular Ignacio de la Calle, for extensive help with the {\sc xmm-esas} package. RJvW acknowledges funding from the Royal Netherlands Academy of Arts and Sciences. MB and MH acknowledge support by the research group FOR 1254 funded by the Deutsche Forschungsgemeinschaft (DFG). AS was supported by Einstein Postdoctoral Fellowship grant number PF9-00070 awarded by the Chandra X-ray Center, which is operated by the Smithsonian Astrophysical Observatory for NASA under contract NAS8-03060. This research is based on data from observations obtained with \xmm\, an ESA science mission with instruments and contributions directly funded by ESA Member States and the USA (NASA).

\bibliographystyle{mn2e}
\bibliography{bibliography}

\appendix

\section{Spectral fits}

\begin{figure}
 \begin{center}
  \includegraphics[width=\columnwidth,keepaspectratio=true,clip=true,trim=0.5cm 0cm 1cm 1.5cm]{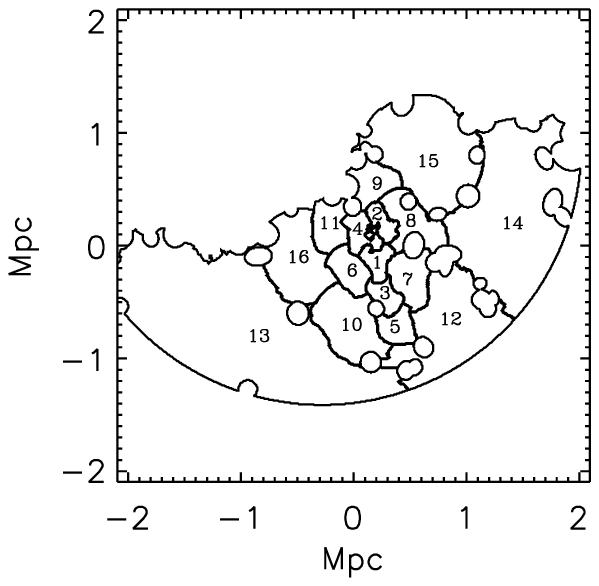}
 \end{center}
 \caption{Bin map used for creating the temperature and pressure maps in Figures \ref{fig:txpmap-freenh} and \ref{fig:txpmap}. Best-fit spectral parameters for each bin are listed in Table \ref{tab:spectralfits}, and the fitted spectra are shown in Figures \ref{fig:binsp-freenh} and \ref{fig:binsp}.}
 \label{fig:binnum}
\end{figure}

\begin{table*}
 \caption{Best-fit parameters to the spectra extracted from the bins in Figure \ref{fig:binnum}. Temperatures are in units of keV, normalizations are in default {\sc Xspec} units per square arcmin, and X-ray column densities are in units of $10^{21}$ cm$^{-2}$. The bins were fitted in parallel. The fit with a fixed $N_{\rm H}$ of $3.22\times 10^{21}$ cm$^{-2}$ had a reduced $\chi^2$ of 1.14. The colums show: (1) bin number, (2) best-fit temperature for $N_{\rm H} = 3.22\times 10^{21}$ cm$^{-2}$, (3) best-fit normalization for $N_{\rm H} = 3.22\times 10^{21}$ cm$^{-2}$, (4) best-fit temperature obtained with a free $N_{\rm H}$, (5) best-fit normalization obtained with a free $N_{\rm H}$, (6) best-fit $N_{\rm H}$.}
\begin{center}
 \begin{tabular}{lccccc}
  \hline
    Bin & $T_{\rm X}$ & $\mathcal{N}_{\rm X}$ & $T_{\rm X}$ & $\mathcal{N}_{\rm X}$ & $N_{\rm H}$\\
  \hline
    1 & $8.3_{-0.54}^{+0.67}$ & $(2.8\pm 0.055) \times 10^{-4}$ & $7.7_{-0.75}^{+0.79}$ & $2.9_{-0.097}^{+0.099} \times 10^{-4}$ & $3.40\pm 0.15$ \\
    2 & $8.0_{-0.56}^{+0.57}$ & $(2.4\pm 0.049) \times 10^{-4}$ & $8.6_{-0.84}^{+1.3}$ & $2.4_{-0.070}^{+0.083} \times 10^{-4}$ & $3.08\pm 0.14$ \\
    3 & $8.3_{-0.55}^{+0.77}$ & $2.9_{-0.057}^{+0.058} \times 10^{-4}$ & $7.1_{-0.59}^{+0.77}$ & $3.1_{-0.10}^{+0.11} \times 10^{-4}$ & $3.57_{-0.15}^{+0.16}$ \\
    4 & $9.6_{-0.90}^{+0.91}$ & $(2.2\pm 0.042) \times 10^{-4}$ & $12.1_{-1.9}^{+2.3}$ & $(2.1\pm 0.044) \times 10^{-4}$ & $2.93_{-0.13}^{+0.14}$ \\
    5 & $8.5_{-0.61}^{+1.1}$ & $2.7_{-0.059}^{+0.058} \times 10^{-4}$ & $10.3_{-1.3}^{+1.9}$ & $2.6_{-0.072}^{+0.079} \times 10^{-4}$ & $2.97_{-0.14}^{+0.15}$ \\
    6 & $8.4_{-0.59}^{+0.96}$ & $(1.8\pm 0.038) \times 10^{-4}$ & $7.3_{-0.63}^{+0.82}$ & $1.9_{-0.066}^{+0.065} \times 10^{-4}$ & $3.53\pm 0.15$ \\
    7 & $8.3_{-0.52}^{+0.62}$ & $(1.7\pm 0.032) \times 10^{-4}$ & $6.8_{-0.48}^{+0.72}$ & $1.9_{0.062}^{+0.066} \times 10^{-4}$ & $3.64\pm 0.14$ \\
    8 & $7.7\pm 0.48$ & $(1.5\pm 0.026) \times 10^{-4}$ & $8.2_{-0.65}^{+0.63}$ & $1.5_{0.037}^{+0.040} \times 10^{-4}$ & $3.13_{-0.10}^{+0.11}$ \\
    9 & $10.5_{-0.94}^{+1.2}$ & $(1.3\pm 0.027) \times 10^{-4}$ & $9.4\pm 1.3$ & $1.4_{-0.042}^{+0.053} \times 10^{-4}$ & $3.40_{-0.14}^{+0.16}$ \\
   10 & $10.0_{-0.97}^{+1.0}$ & $(1.0\pm 0.020) \times 10^{-4}$ & $9.4_{-1.2}^{+1.3}$ & $1.0_{-0.029}^{+0.033} \times 10^{-4}$ & $3.33_{-0.12}^{+0.13}$ \\
   11 & $11.6_{-1.1}^{1.5}$ & $(1.1\pm 0.022) \times 10^{-4}$ & $7.7_{-0.75}^{+0.78}$ & $1.3_{-0.041}^{+0.042} \times 10^{-4}$ & $3.86_{-0.13}^{+0.14}$ \\
   12 & $7.2_{-0.57}^{+0.69}$ & $(5.2\pm 0.12) \times 10^{-5}$ & $7.4_{-0.74}^{+0.87}$ & $(5.2\pm 0.17) \times 10^{-5}$ & $3.23_{-0.093}^{+0.095}$ \\
   13 & $6.8_{-1.8}^{+4.1}$ & $5.5_{-0.55}^{+0.66} \times 10^{-6}$ & $4.3_{-0.94}^{+0.99}$ & $7.6_{-0.81}^{+1.1} \times 10^{-6}$ & $3.65_{-0.089}^{+0.10}$ \\
   14 & $5.2_{-0.94}^{+1.8}$ & $9.9_{-0.87}^{+0.83} \times 10^{-6}$ & $6.2_{-1.6}^{+2.1}$ & $9.6_{-0.86}^{+1.1} \times 10^{-6}$ & $3.24_{-0.087}^{+0.095}$ \\
   15 & $11.3_{-1.2}^{+2.6}$ & $2.3_{-0.072}^{+0.074} \times 10^{-5}$ & $9.3_{-1.6}^{+1.8}$ & $2.5_{-0.093}^{+0.11} \times 10^{-5}$ & $3.52_{-0.091}^{+0.098}$ \\
   16 & $13.0_{-2.0}^{+2.4}$ & $(4.7\pm 0.11) \times 10^{-5}$ & $12.2_{-1.9}^{+2.6}$ & $(4.8\pm 0.12) \times 10^{-5}$ & $3.33_{-0.093}^{+0.096}$ \\
  \hline
 \end{tabular}
\end{center}
 \label{tab:spectralfits}
\end{table*}

\begin{figure*}
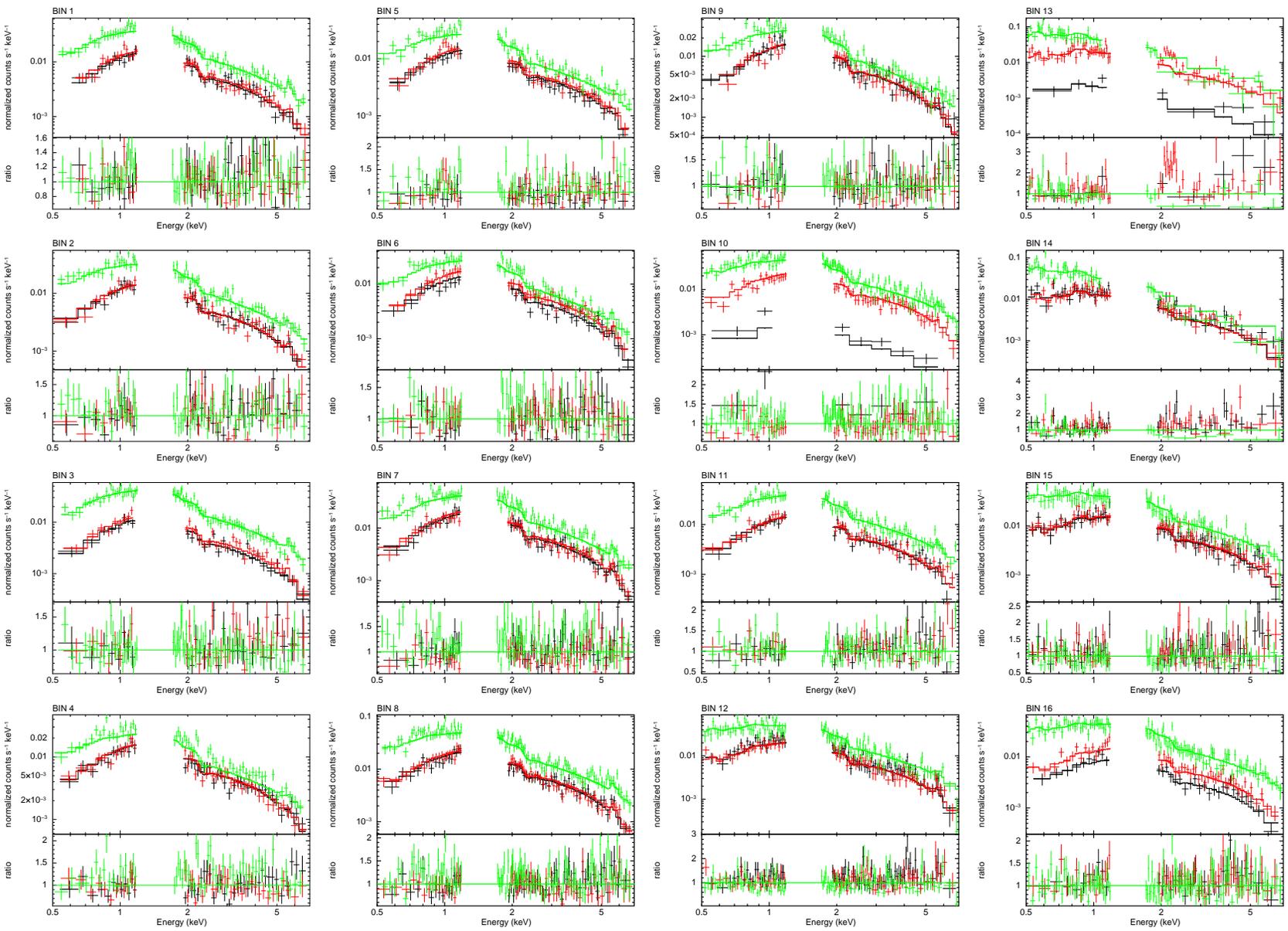

 \begin{center}
  \includegraphics[width=0.22\textwidth]{bin20-freenh-sp.ps}
  \includegraphics[width=0.22\textwidth]{bin21-freenh-sp.ps}
  \includegraphics[width=0.22\textwidth]{bin25-freenh-sp.ps}
  \includegraphics[width=0.22\textwidth]{bin26-freenh-sp.ps}
  \includegraphics[width=0.22\textwidth]{bin13-freenh-sp.ps}
  \includegraphics[width=0.22\textwidth]{bin15-freenh-sp.ps}
  \includegraphics[width=0.22\textwidth]{bin17-freenh-sp.ps}
  \includegraphics[width=0.22\textwidth]{bin19-freenh-sp.ps}
  \includegraphics[width=0.22\textwidth]{bin5-freenh-sp.ps}
  \includegraphics[width=0.22\textwidth]{bin7-freenh-sp.ps}
  \includegraphics[width=0.22\textwidth]{bin8-freenh-sp.ps}
  \includegraphics[width=0.22\textwidth]{bin9-freenh-sp.ps}
  \includegraphics[width=0.22\textwidth]{bin0-freenh-sp.ps}
  \includegraphics[width=0.22\textwidth]{bin1-freenh-sp.ps}
  \includegraphics[width=0.22\textwidth]{bin2-freenh-sp.ps}
  \includegraphics[width=0.22\textwidth]{bin4-freenh-sp.ps}
 \end{center}
 \caption{Fits to the spectra extracted from the bins in Figure \ref{fig:binnum}, obtained with a free X-ray column density. MOS1 and MOS2 data are plotted in black and red, respectively, while pn is plotted in green. The bottom pannel shows the ratio of the data to the model.}
 \label{fig:binsp-freenh}
\end{figure*}

\section{Results with fixed $N_{\rm H}$}

\begin{table}
 \caption{Fitted background and foreground parameters, obtained using a fixed X-ray column density of $N_{\rm H}=3.22\times 10^{21}$ cm$^{-2}$.}
\begin{center}
 \begin{tabular}{lccc}
  \hline
   Component & $\Gamma$ & $T_{\rm X}$ & $N_{\rm X}$ \\
  \hline
    LHB & -- & $0.08^\dagger$ & $(3.4\pm 0.29) \times 10^{-7}$  \\
    GH  & -- & $0.14_{-0.0025}^{+0.0029}$ & $1.4_{-0.083}^{+0.081} \times 10^{-5}$ \\
    HF  & -- & $0.62_{-0.022}^{+0.024}$ & $8.4_{-0.58}^{+0.54} \times 10^{-7}$ \\
    CXB & $1.41^\dagger$ & -- & $1.1_{-0.044}^{+0.045} \times 10^{-6}$ \\
  \hline
 \end{tabular}
\end{center}
 \hspace{-0cm}$\dagger$ {\footnotesize frozen}
 \label{tab:bkgmod}
\end{table}

\begin{figure*}
 \begin{center}
  \includegraphics[width=0.49\textwidth,keepaspectratio=true,clip=true,trim=0.2cm 0cm 1.3cm 0cm]{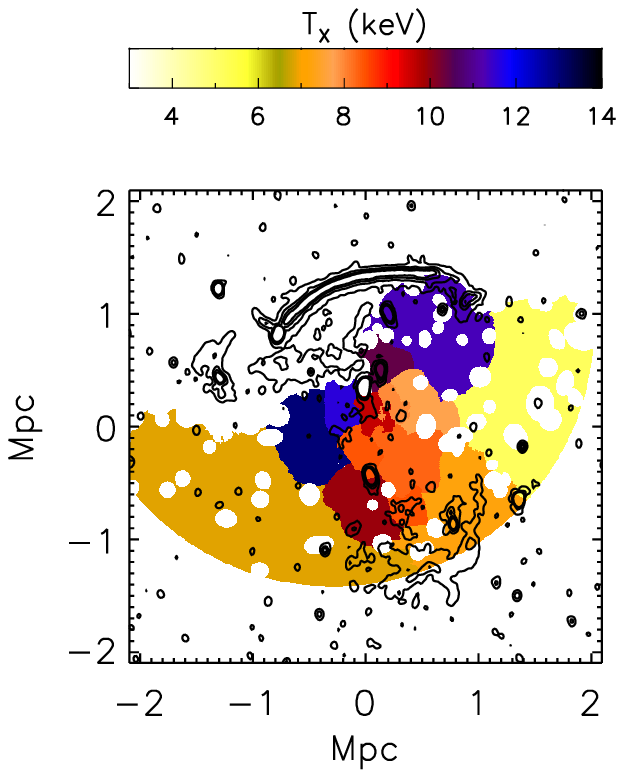}
  \includegraphics[width=0.49\textwidth,keepaspectratio=true,clip=true,trim=0.2cm 0cm 1.3cm 0cm]{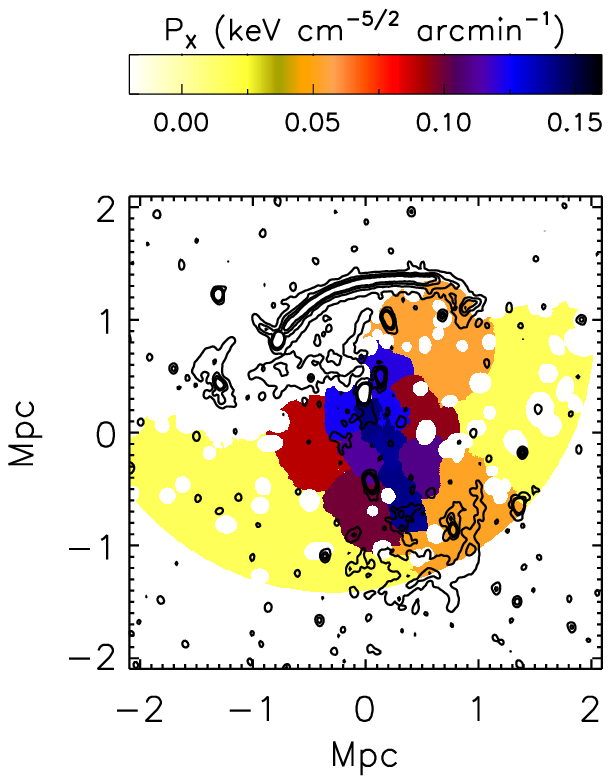}
 \end{center}
 \caption{Temperature and pseudo-pressure maps of the cluster, obtained using a fixed X-ray column density of $N_{\rm H}=0.322\times 10^{21}$ cm$^{-2}$. Spectra were extracted from bins that follow the surface brightness contours, and have a total of $\sim 3600$ counts after the subtraction of the instrumental background. The pressure was calculated as the product between the temperature and the square root of the normalization of the spectral component describing ICM emission. Overlaid are the radio contours also shown in Figure \ref{fig:xmm-radio}.}
 \label{fig:txpmap}
\end{figure*}

\begin{figure*}
 \begin{center}
  \includegraphics[width=0.22\textwidth]{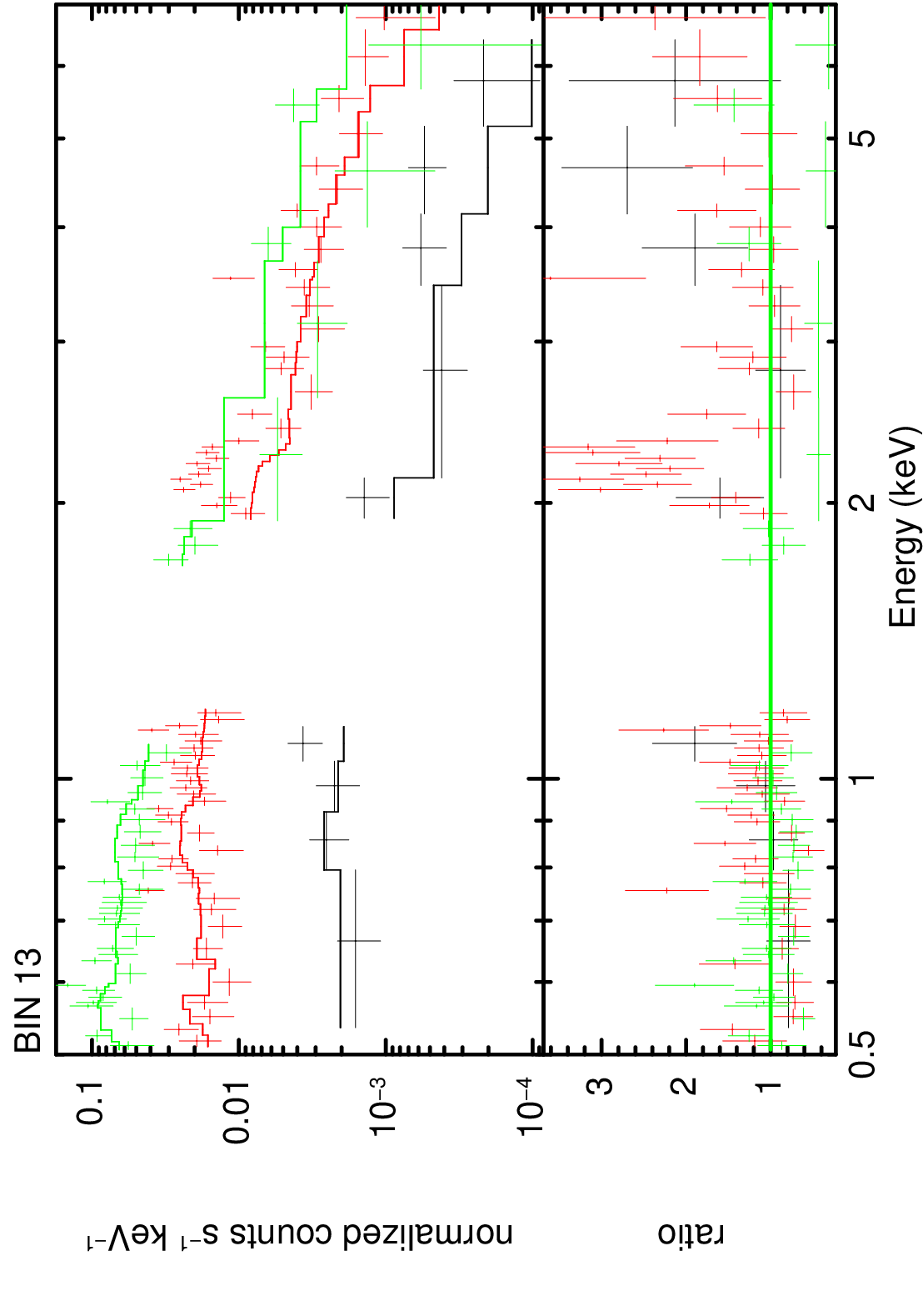}
  \includegraphics[width=0.22\textwidth]{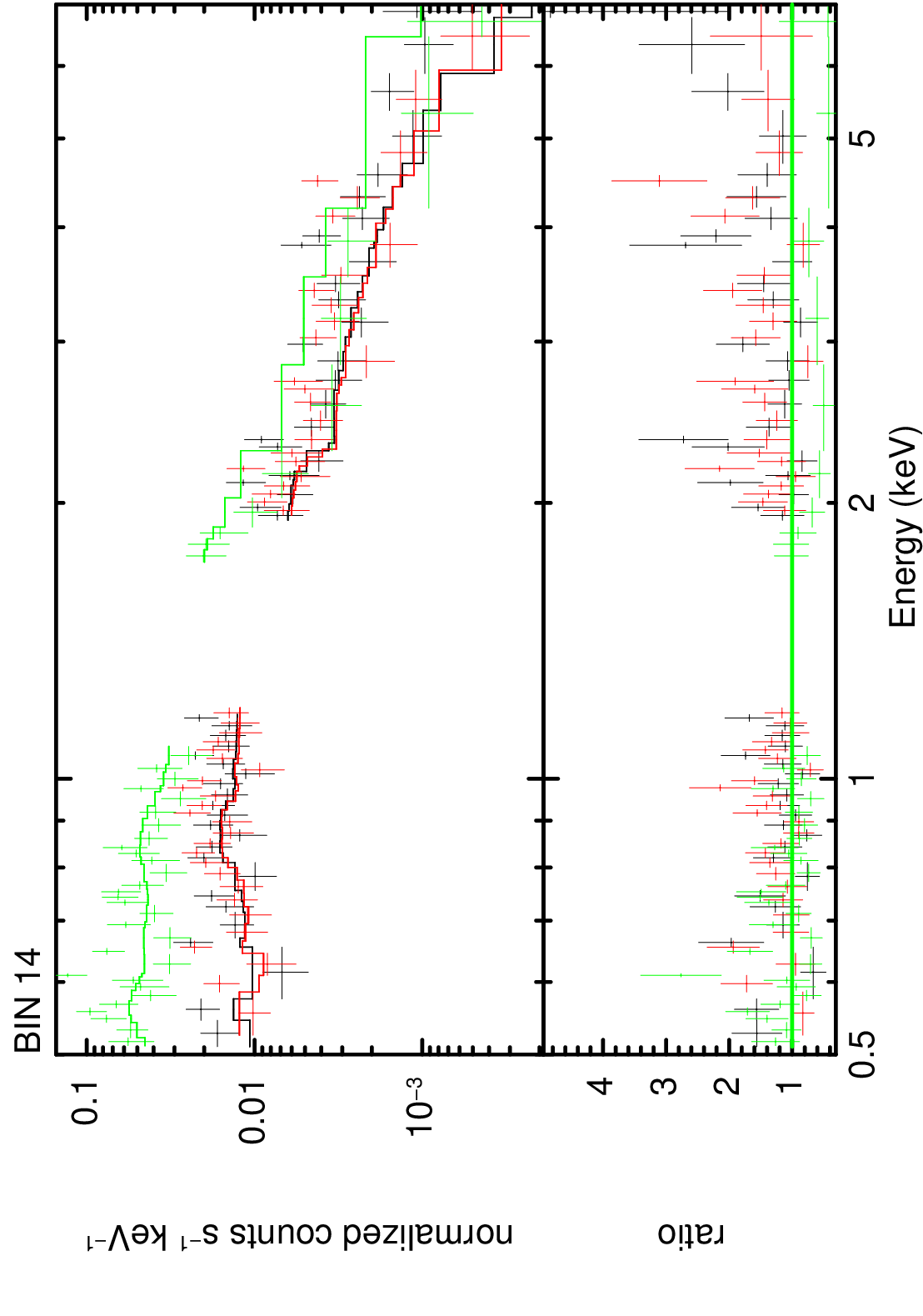}
  \includegraphics[width=0.22\textwidth]{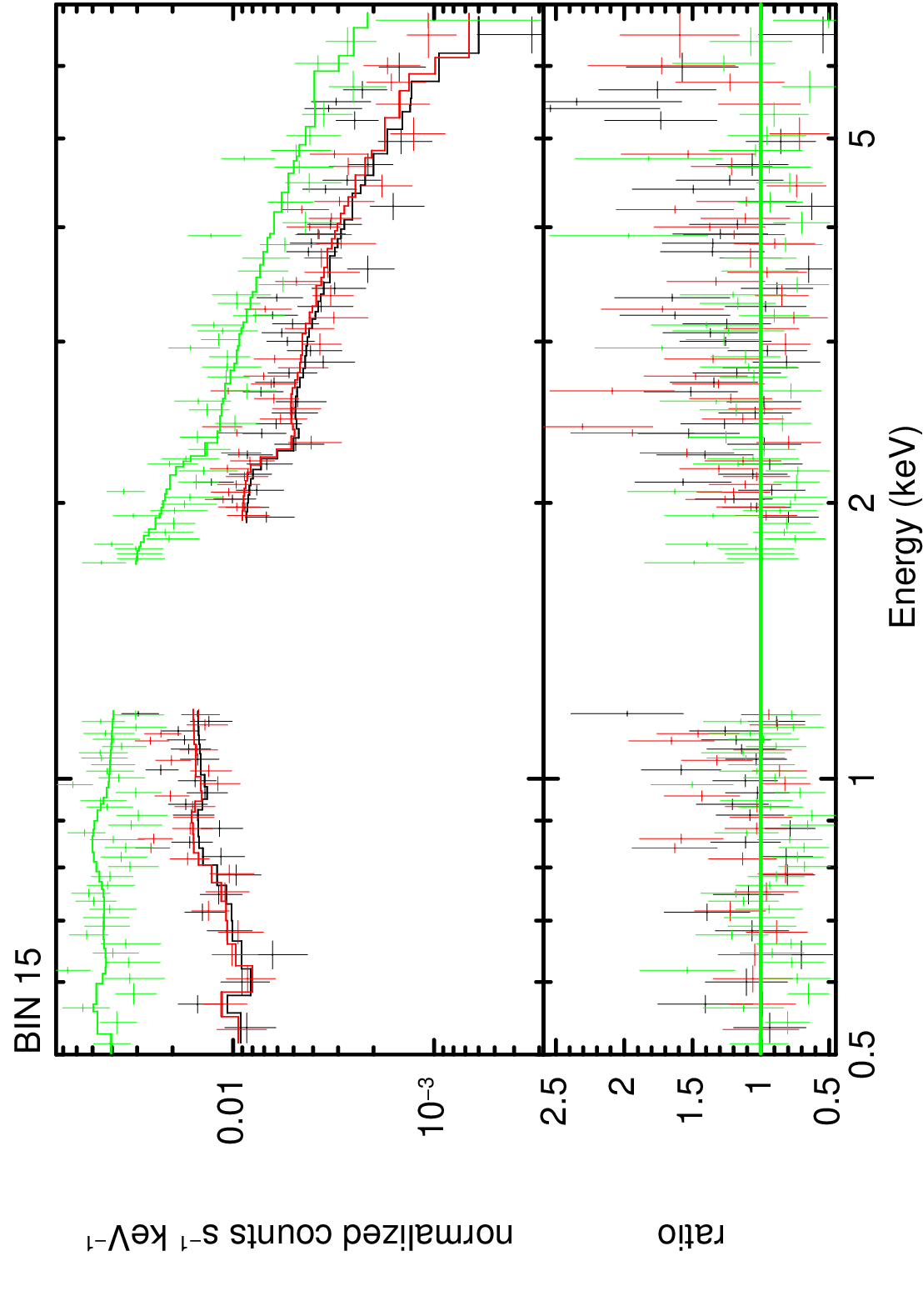}
  \includegraphics[width=0.22\textwidth]{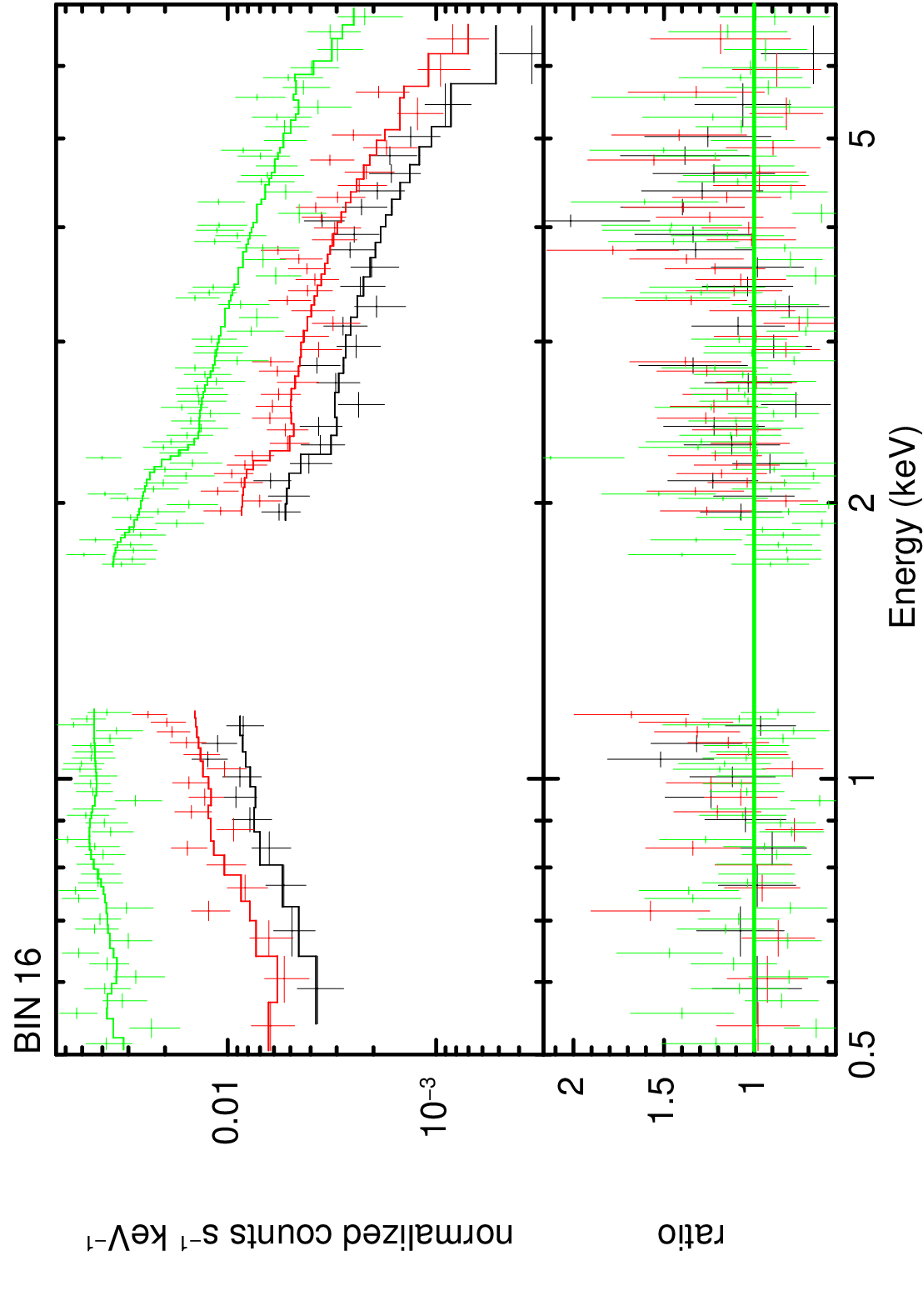}
  \includegraphics[width=0.22\textwidth]{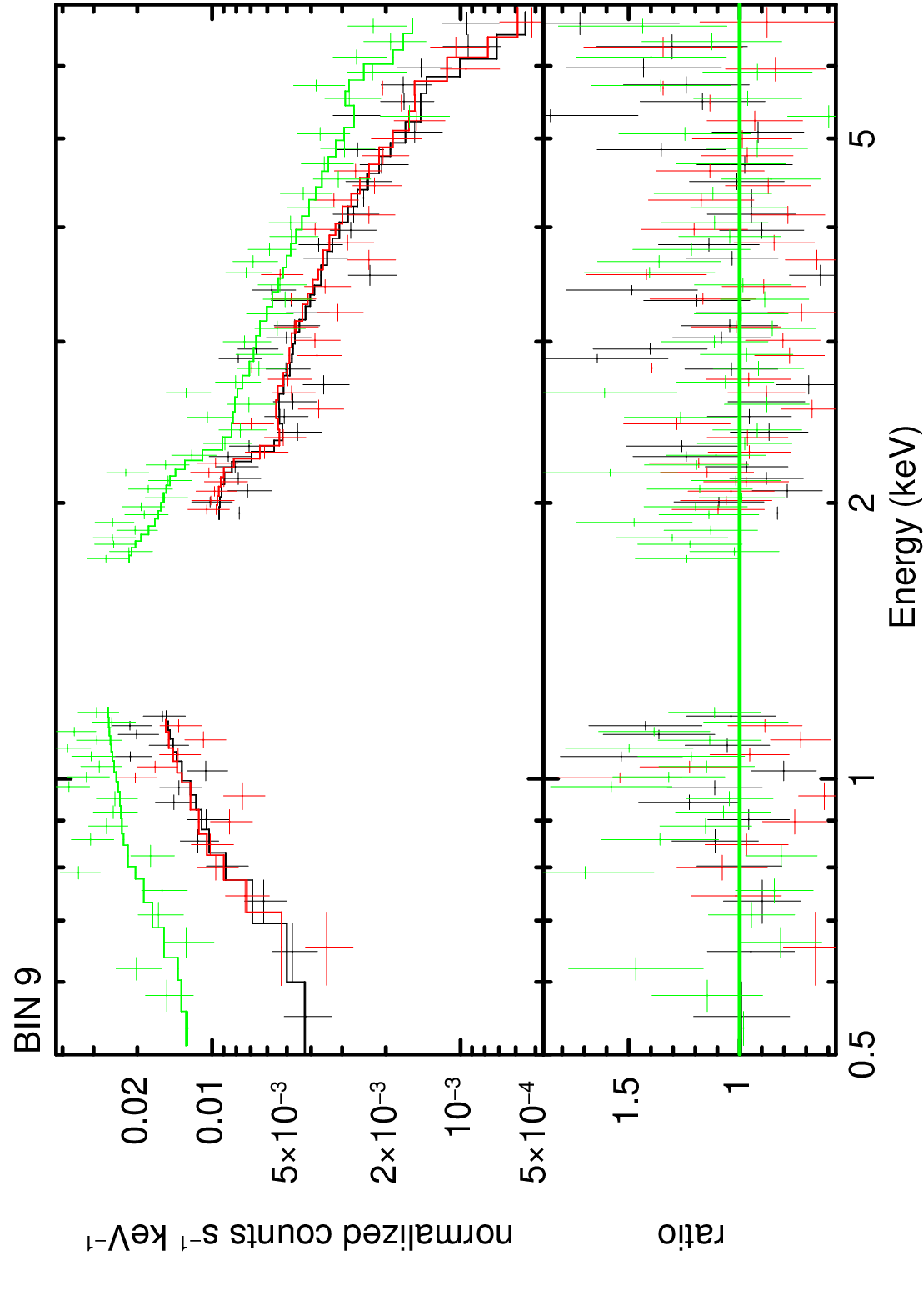}
  \includegraphics[width=0.22\textwidth]{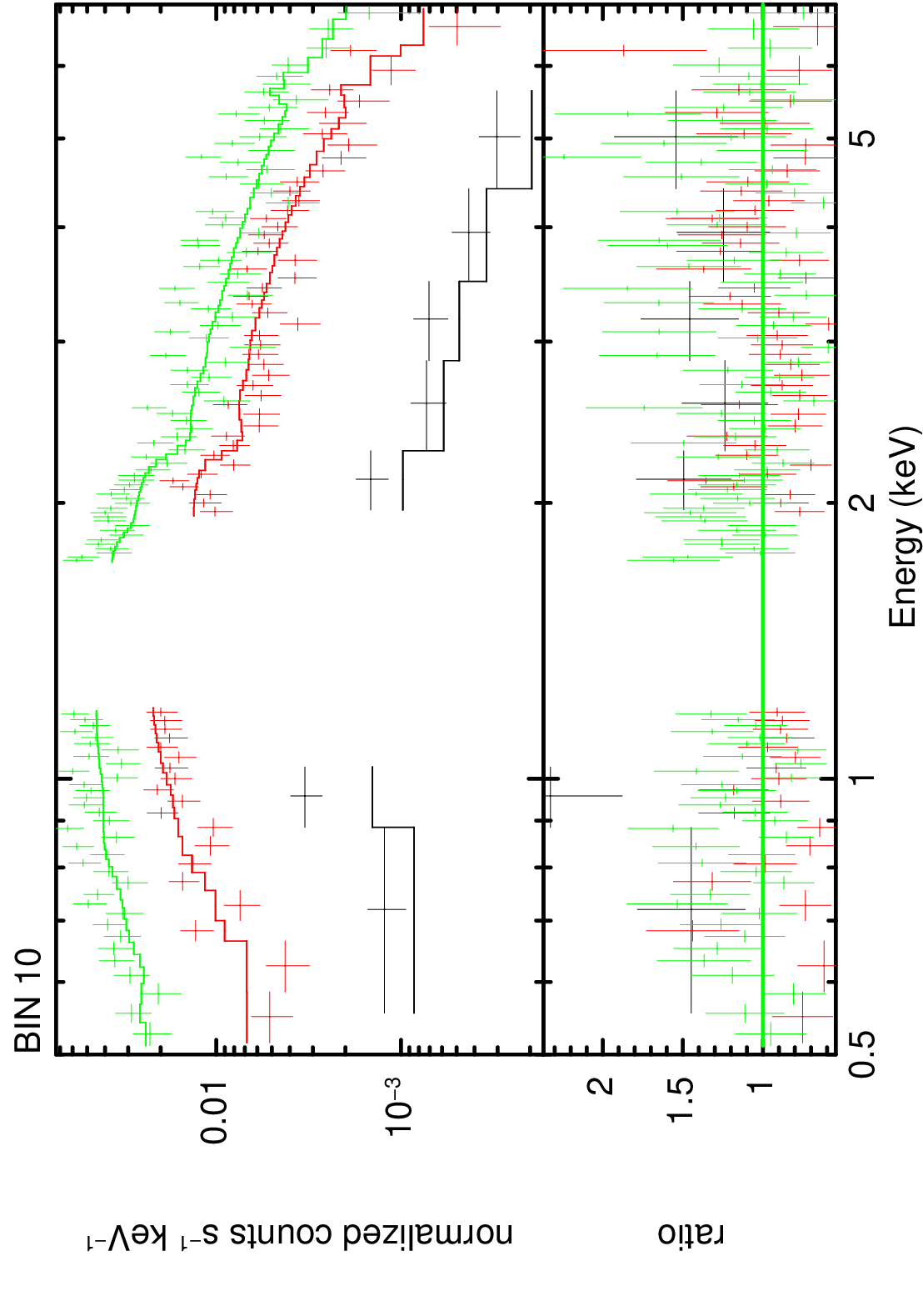}
  \includegraphics[width=0.22\textwidth]{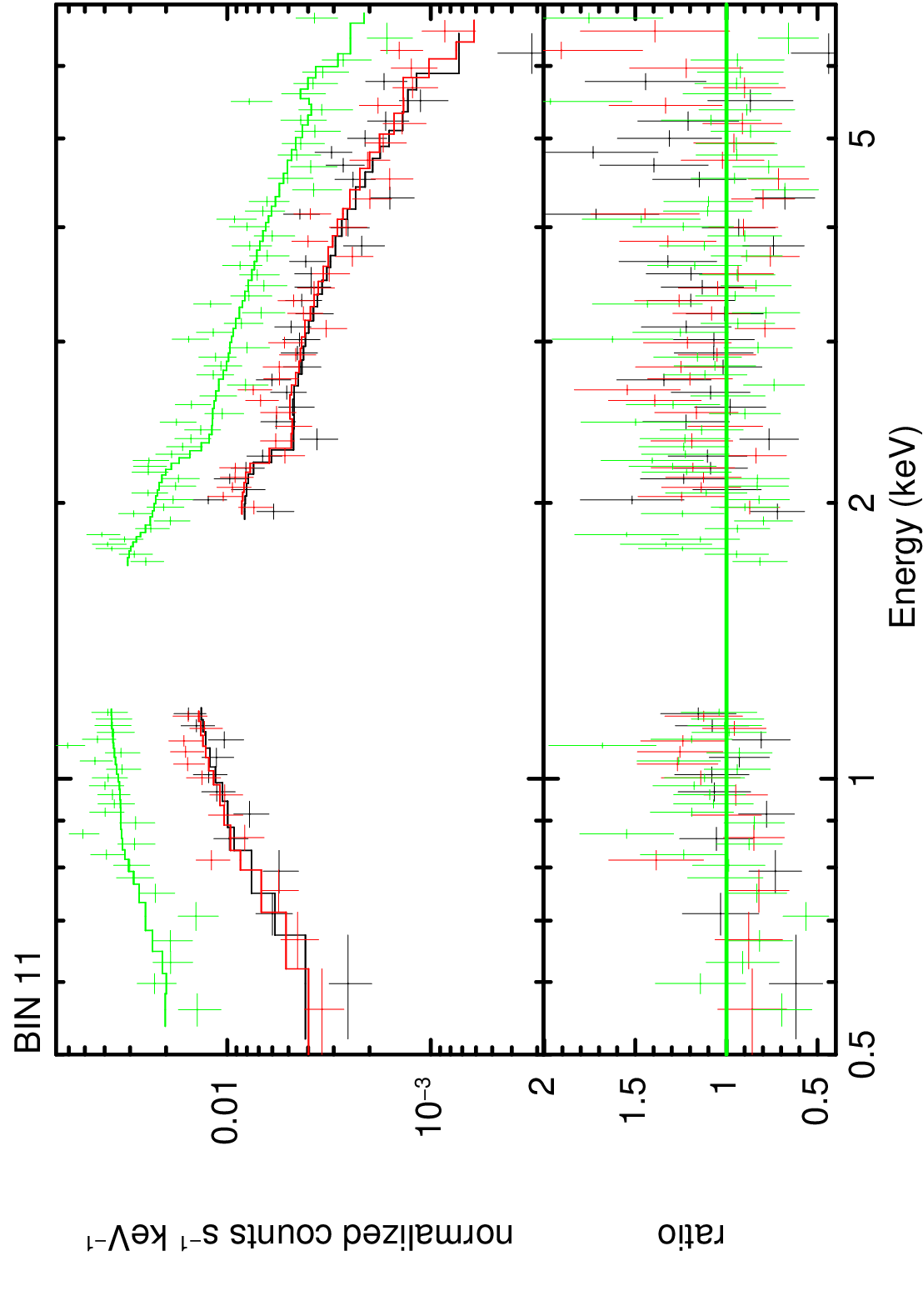}
  \includegraphics[width=0.22\textwidth]{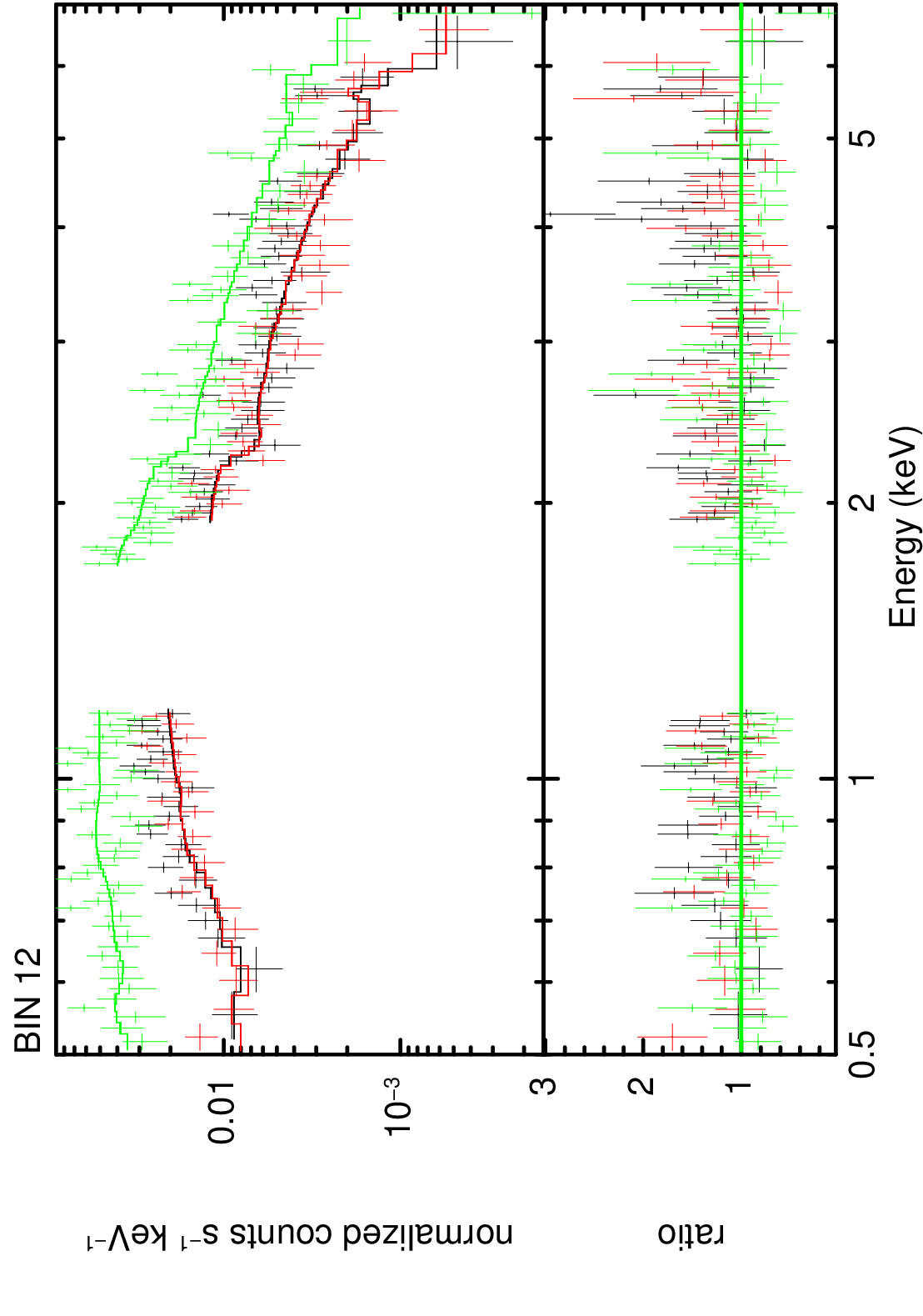}
  \includegraphics[width=0.22\textwidth]{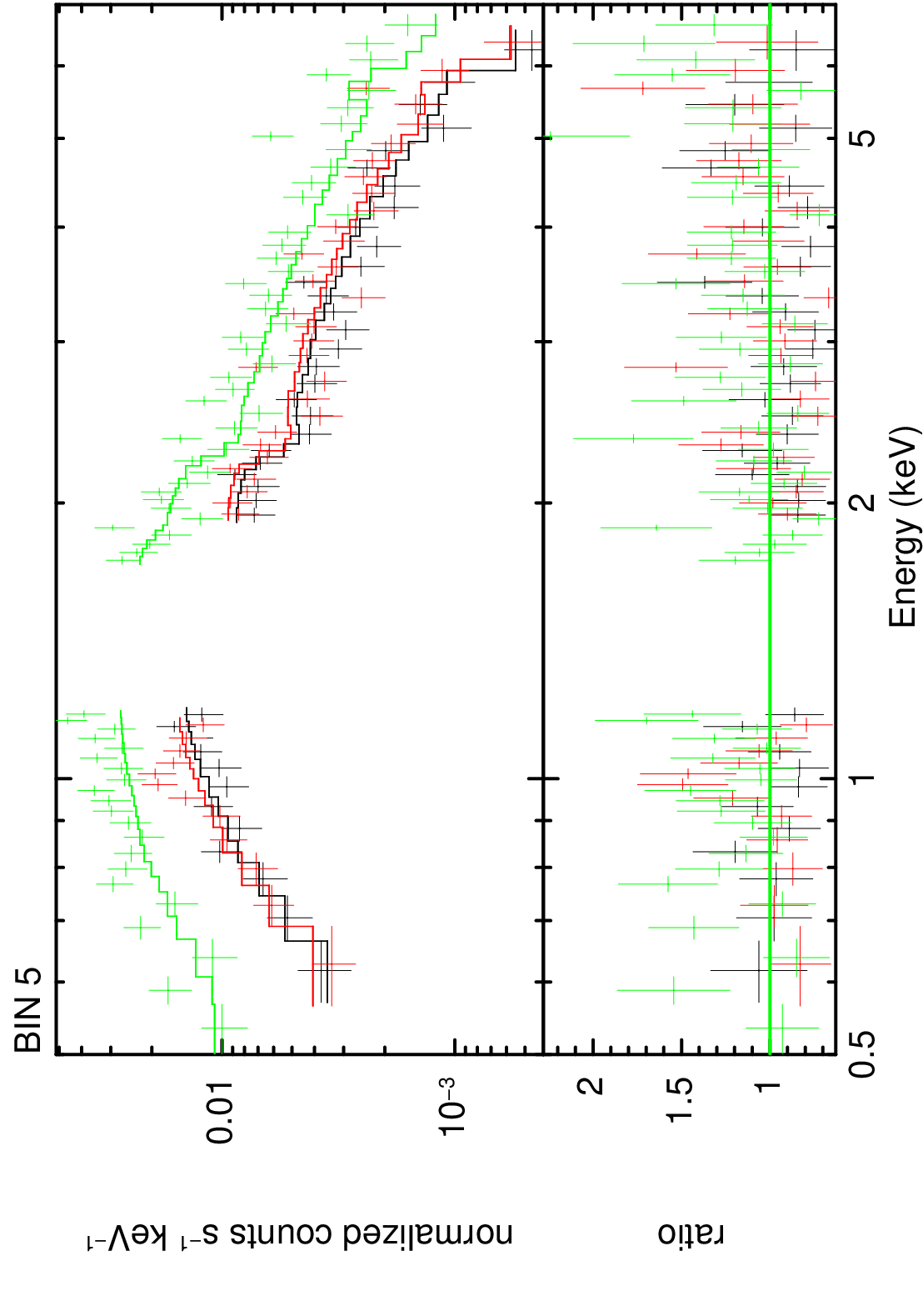}
  \includegraphics[width=0.22\textwidth]{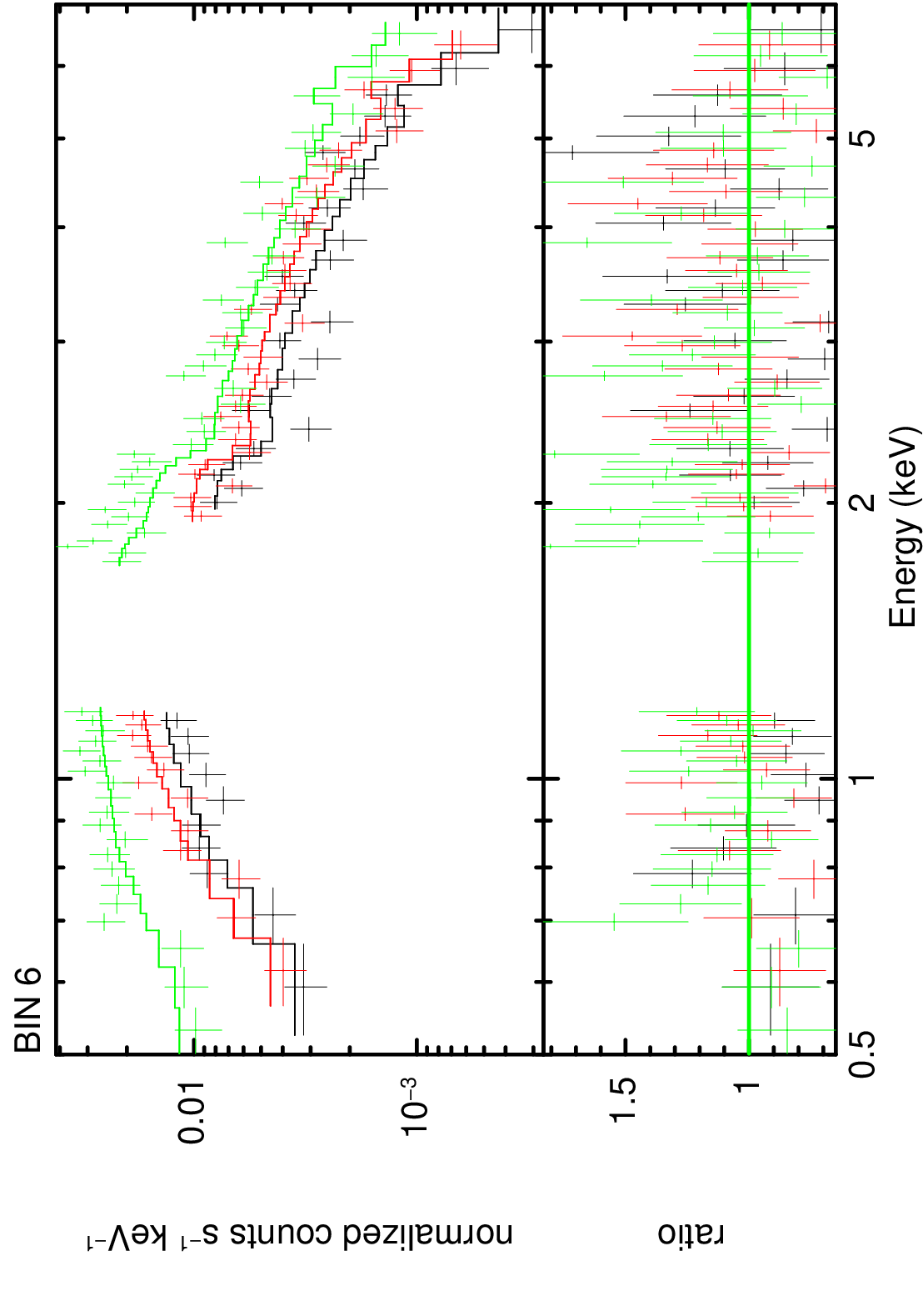}
  \includegraphics[width=0.22\textwidth]{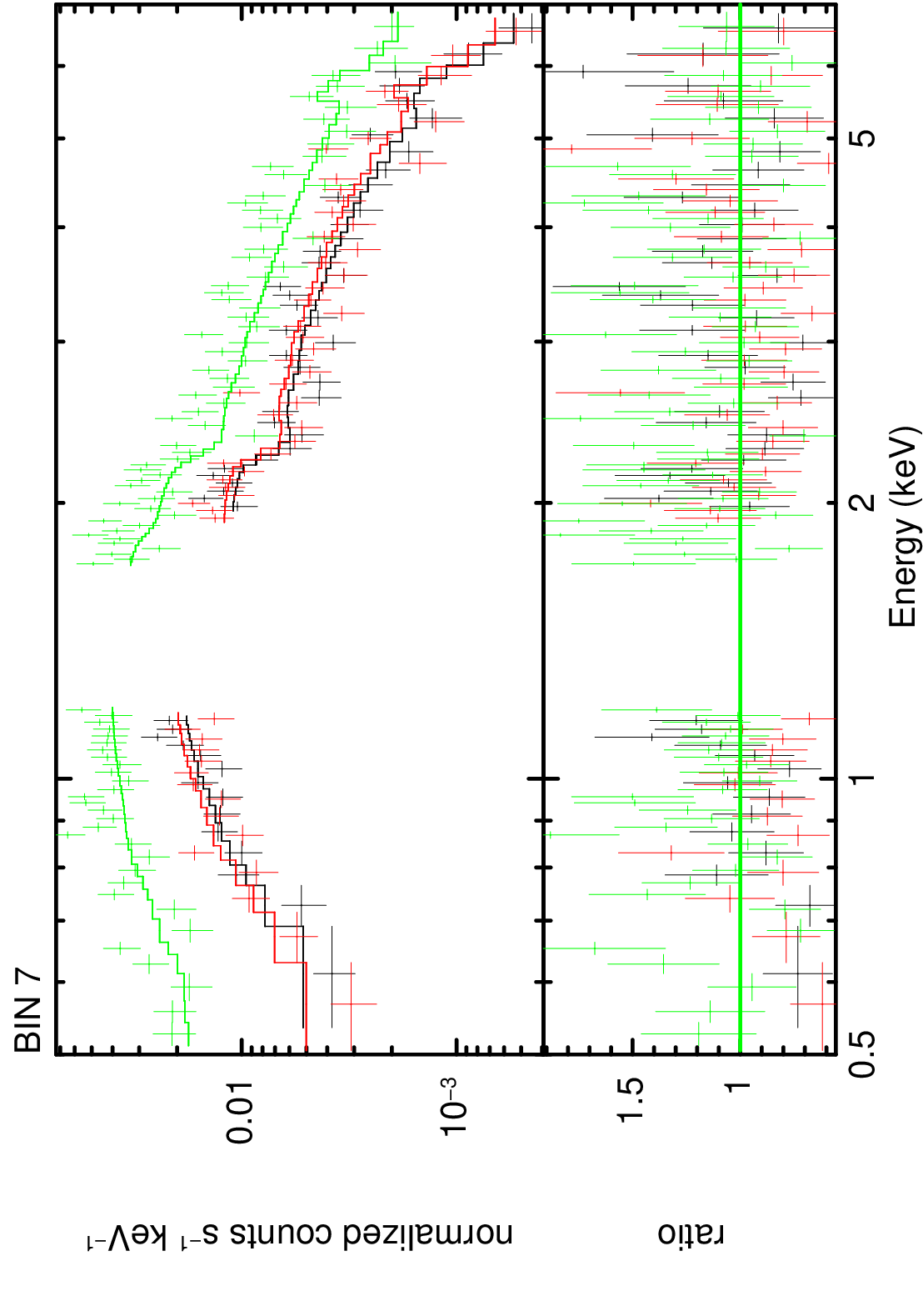}
  \includegraphics[width=0.22\textwidth]{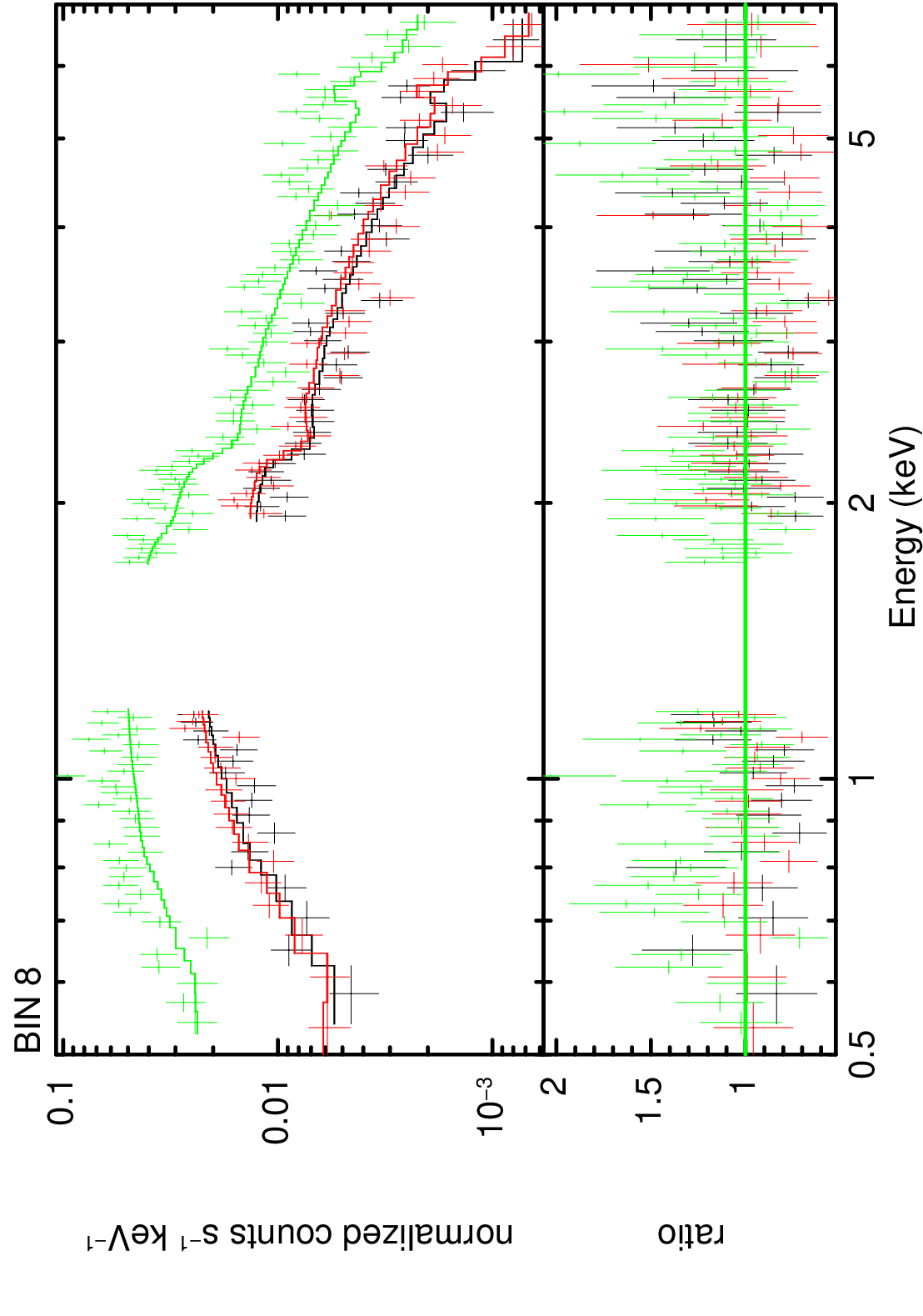}
  \includegraphics[width=0.22\textwidth]{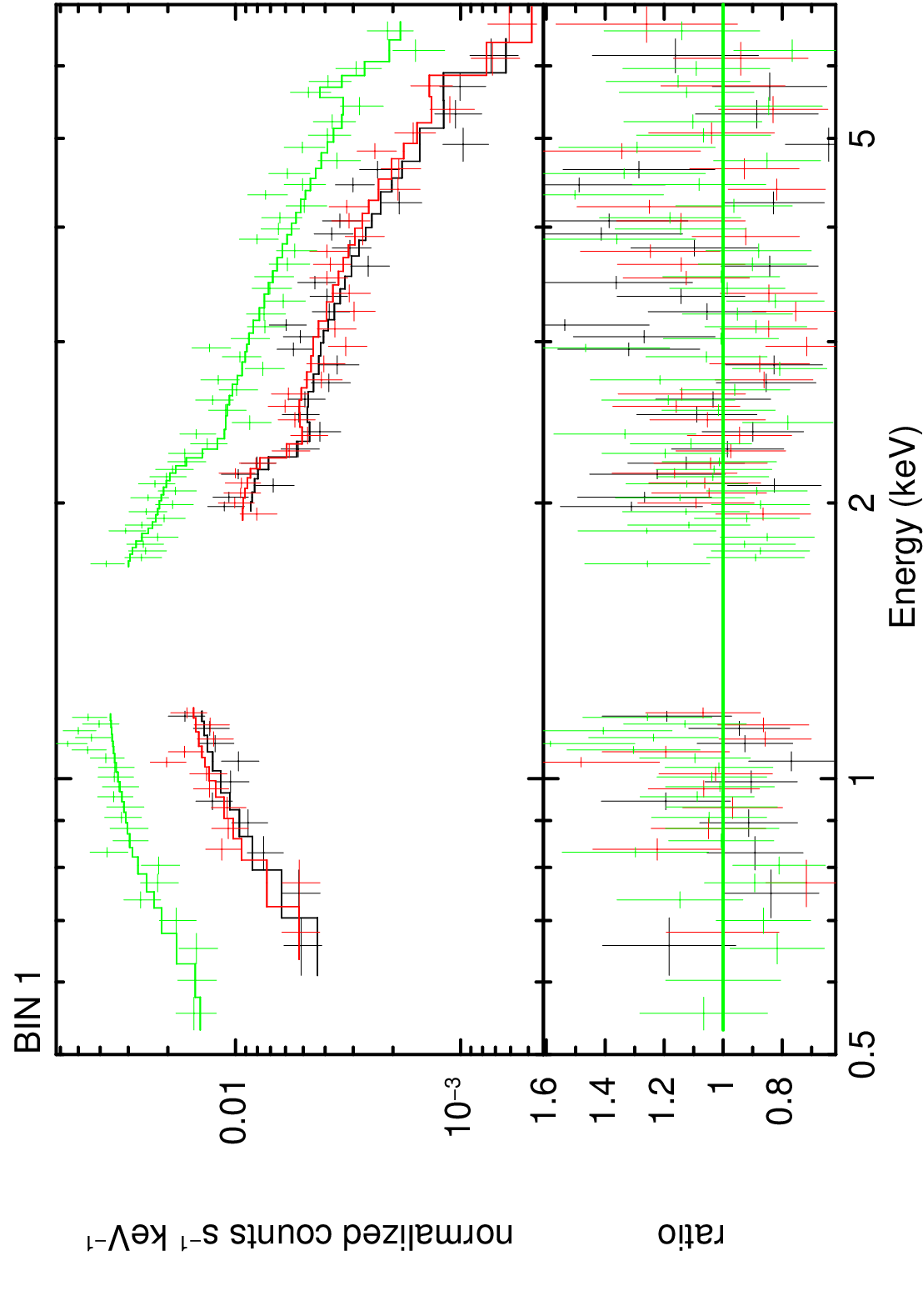}
  \includegraphics[width=0.22\textwidth]{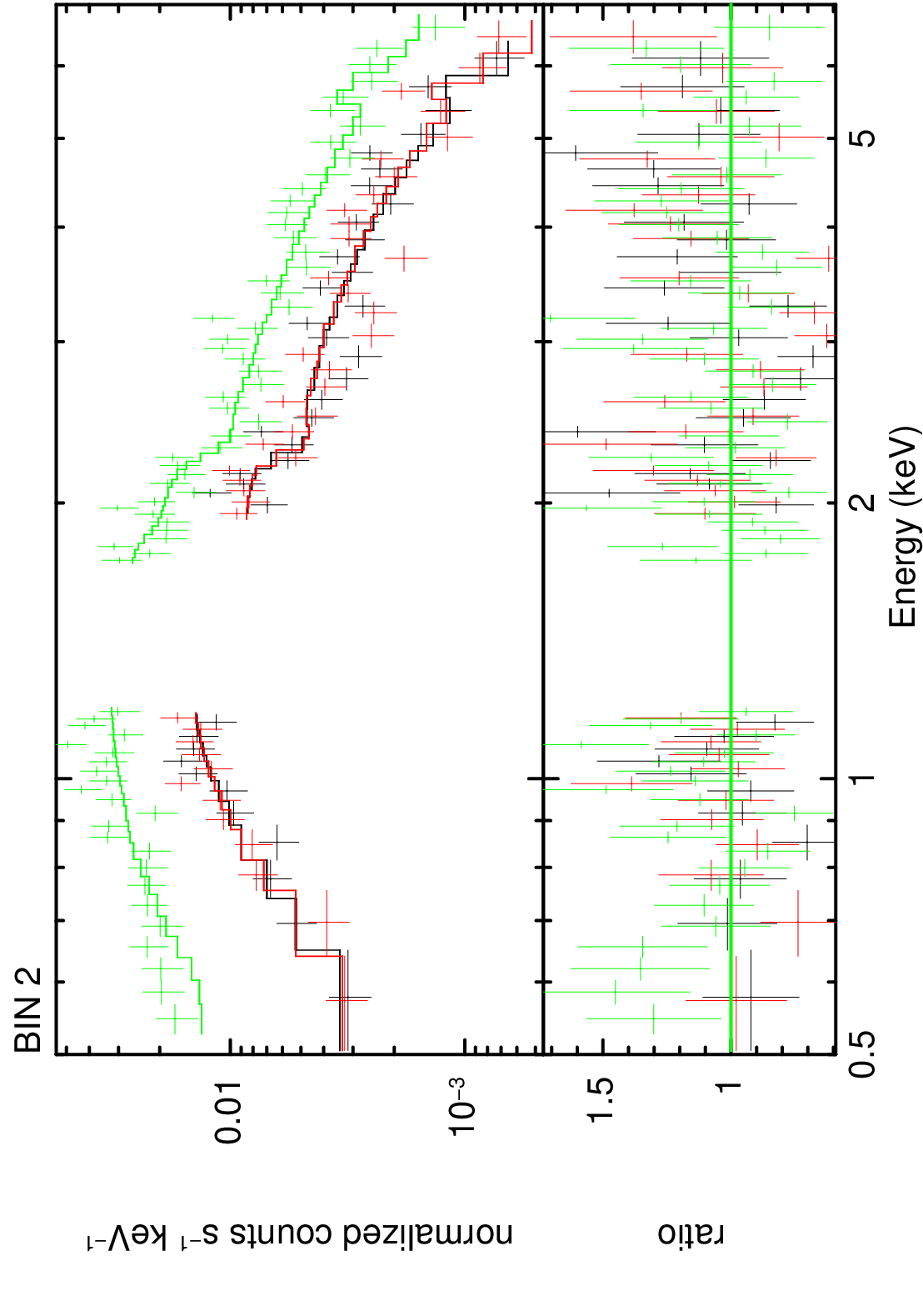}
  \includegraphics[width=0.22\textwidth]{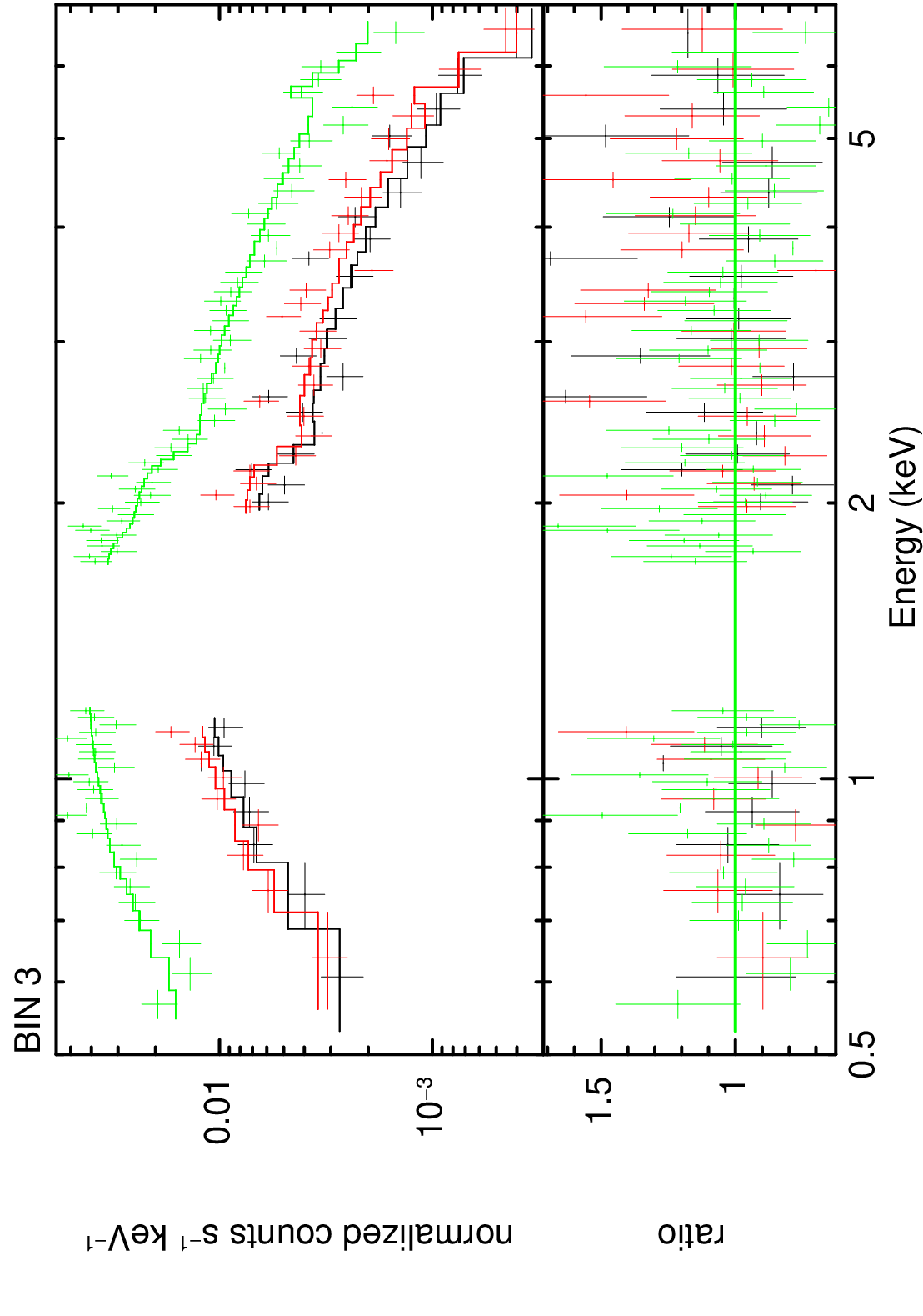}
  \includegraphics[width=0.22\textwidth]{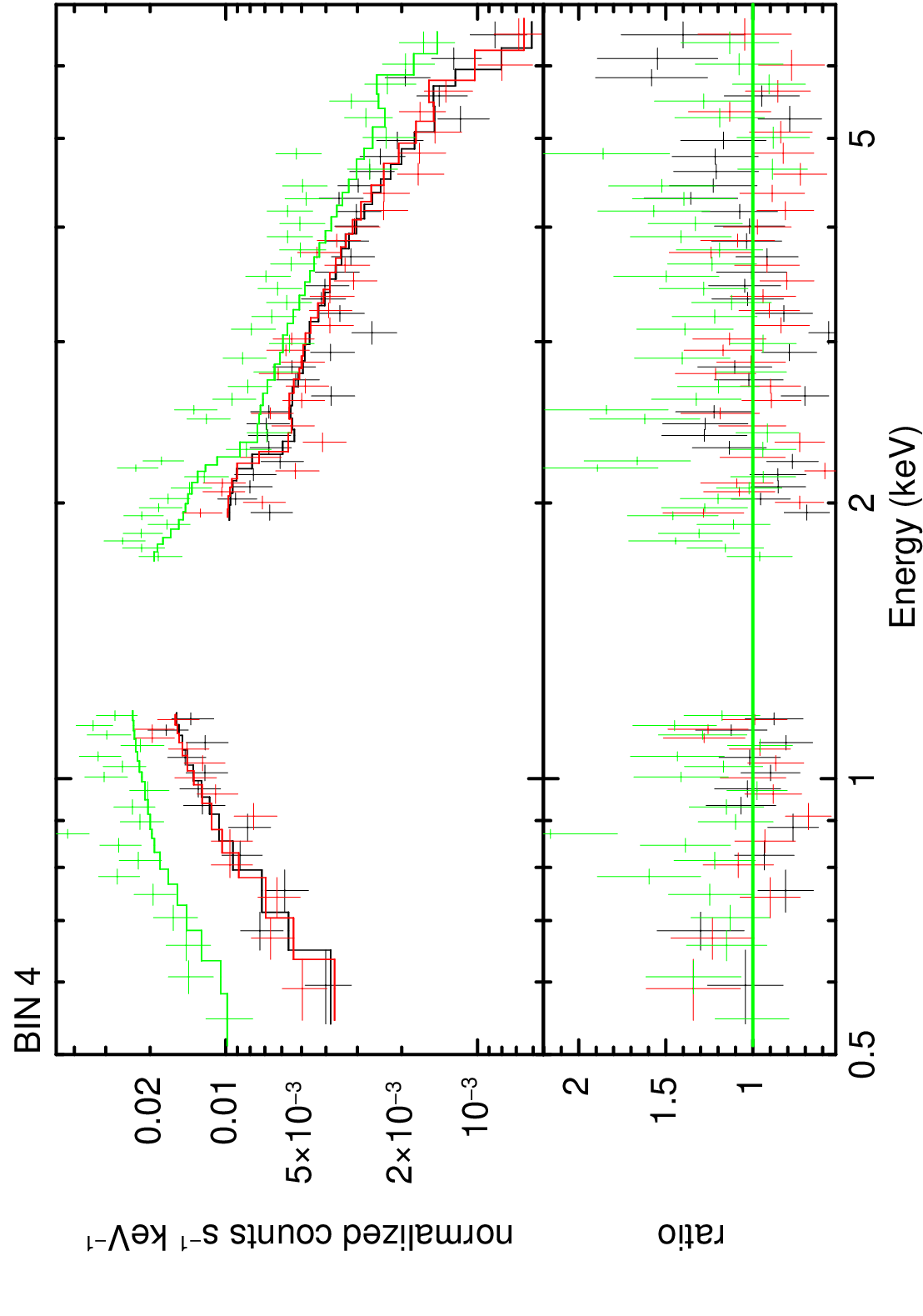}
 \end{center}
 \caption{Fits to the spectra extracted from the bins in Figure \ref{fig:binnum}, obtained using a fixed X-ray column density of $3.22\times 10^{21}$ cm$^{-2}$. MOS1 and MOS2 data are plotted in black and red, respectively, while pn is plotted in green. The bottom pannel shows the ratio of the data to the model.}
 \label{fig:binsp}
\end{figure*}

\label{lastpage}

\end{document}